\documentclass[onecolumn,sort&compress,numbers]{els-mrw} 

\usepackage{amsmath,amssymb,amsfonts,amsthm,makeidx,graphicx}
\usepackage{txfonts}
\usepackage{helvet}
\usepackage{dsfont}
\usepackage{slashed}
\usepackage{setspace}

\usepackage[colorlinks=true,linkcolor=blue,citecolor=blue,urlcolor=blue]{hyperref}

\definecolor{myblue}{RGB}{0,0,255}
\definecolor{myred}{RGB}{204,0,0}

\newcommand{\checked}{\textcolor{myblue}{\ding{51}}}
\newcommand{\unchecked}{\textcolor{myred}{\ding{55}}}

\def\bea#1\eea{\begin{align}#1\end{align}}

\newcommand{\bef}{\begin{figure}[h!tb]\centering}
\newcommand{\eef}{\end{figure}}

\usepackage[table]{xcolor}
\usepackage{pifont}
\usepackage{amsmath}
\usepackage{array} 
\usepackage{newtxtext}
\usepackage{newtxmath}

\begin{document}


\chapter{Parton Fragmentation Functions}\label{chap1}

\author[1]{Hongxi Xing}%

\address[1]{\orgname{South China Normal University}, \orgdiv{Institute of Quantum Matter}, \orgaddress{Guangzhou, China}}


\maketitle

\begin{abstract}[Abstract]
This chapter provides a pedagogical and comprehensive introduction to parton fragmentation functions (FFs), the essential non-perturbative quantities in Quantum Chromodynamics (QCD) that describe the universal process of hadronization, the transition of quarks and gluons into observable hadrons. Formally defined as gauge-invariant matrix elements, the determination of FFs is achieved through global QCD analyses that synthesize world data from three primary processes: electron-positron annihilation, semi-inclusive deep inelastic scattering, and proton-proton collisions. This chapter covers the theoretical foundations of FFs, including their operator definitions and properties, the experimental probes used to constrain them, and the methodologies of modern global analyses. The discussion is extended to more advanced topics, including dihadron FFs, the three-dimensional picture of hadronization provided by spin- and transverse-momentum-dependent FFs, and the modification of fragmentation in a nuclear medium, which is key to probe their fundamental properties.

\end{abstract}

\begin{keywords}
	Parton fragmentation functions\sep QCD evolution equations\sep hadronization \sep high energy scattering \sep polarization
\end{keywords}

\section*{Objectives}
\begin{itemize}
	\item The operator definition of parton fragmentation functions in QCD and their physical interpretation.
	\item The concept of DGLAP evolution for FFs and its role in interpreting experimental data.
        \item  The experimental processes (SIA, SIDIS, pp) that provide constraints on FFs and the structure of their factorized cross sections.
	\item  The methodology of modern global QCD analyses, including parametrization strategies and uncertainty estimation.
        \item  The theory and phenomenology of extended FF topics: dihadron, polarized, and transverse-momentum-dependent FFs, and fragmentation in nuclear medium.
\end{itemize}

\section{Introduction}
\label{sec:introFFs}

In the Standard Model of particle physics, fragmentation functions (FFs) are essential non-perturbative quantities within the theory of quantum chromodynamics (QCD). They describe the process of hadronization, the transition of high-energy quarks and gluons, collectively known as partons, into the composite and color-neutral particles called hadrons that are detected in experiments. This transformation is a direct consequence of color confinement, a key feature of QCD that prohibits the existence of free, colored partons.

In high-energy collisions, such as lepton-hadron deep-inelastic scattering (DIS), electron-positron annihilation ($e^+e^-$), and proton-proton ($pp$) interactions, the energetic partons produced in hard scattering processes evolve into jets of observable hadrons. While the hard scattering is calculable using perturbative QCD, the subsequent hadronization is a complex, long-distance phenomenon. FFs systematically encode this non-perturbative process, serving as a critical component of the QCD factorization framework, which separates the calculable short-distance physics from the universal, long-distance dynamics of hadron formation.

The development of FFs parallels with the growth of QCD itself. The concept first emerged in the early 1970s \cite{Berman:1971xz} with the advent of the parton model. This new ingredient was initially called parton decay functions, and was introduced as a counter part of parton distribution functions (PDFs) to calculate inclusive processes for hadron production at high transverse momentum, including both photonic, leptonic and hadronic scatterings. A crucial first step towards quantifying the hadronization process was the phenomenological model of Feynman and Field \cite{Field:1977fa} in the late 1970s, which described hadronization as a recursive, scale-invariant cascade. This provided an intuitive picture and a means to parameterize experimental data. Concurrently, a seminal theoretical improvement came with the development of the Dokshitzer–Gribov–Lipatov–Altarelli–Parisi (DGLAP) evolution equations \cite{Gribov:1972ri,Gribov:1972rt,Lipatov:1974qm,Dokshitzer:1977sg,Altarelli:1977zs}. These equations demonstrated that while the FFs themselves are non-perturbative, their dependence on the energy scale is universal and can be calculated in perturbative QCD. The theoretical foundation was established in the early 1980s through the pioneering work of Collins and Soper \cite{Collins:1981uw}, who established the rigorous operator definition of FFs as gauge-invariant matrix elements. This work formally proved the universality of FFs, emphasizing their role as fundamental quantities in QCD factorization theorems and enabling the modern era of precision global analyses.

Due to the time-dependent nature and the complexity of out states in the operator definition of FFs, direct calculation of FFs in a first-principle fashion is still in lack of exploration, in sharp contrast with their counter part PDFs, where the relevant lattice QCD investigation has formed a main branch in particle physics community \cite{Lin:2017snn}. QCD global analysis stands as the only reliable quantification of FFs, this interdisciplinary endeavor synthesizes advances across experimental measurements collected from decades of experiments at facilities worldwide, high-precision perturbative QCD calculations, and computational science connecting theory and experiment, which allows for the determination of a universal set of fragmentation functions. 

The framework of fragmentation functions have also been extended beyond the simple one-dimensional, unpolarized case. To capture the full complexity of hadronization, the formalism is generalized to include parton spin and transverse momentum (see Refs. \cite{Metz:2016swz,Albino:2008gy,Boussarie:2023izj} for introduction and reviews). Polarized FFs describe the transfer of spin from a fragmenting parton to a final-state hadron, while transverse momentum dependent (TMD) FFs, such as the well-known Collins function \cite{Collins:1992kk}, provide a three-dimensional picture of hadronization by retaining information on the parton's transverse motion. Furthermore, the concept of dihadron FFs, describing the production of a correlated hadron pair, offers an alternative window into the spin and momentum structure of fragmentation. The study of FFs is also extended to hadronization within a nuclear environment, leading to medium-modified FFs. These are key ingredients for understanding phenomena like jet quenching in the hot, dense quark-gluon plasma formed in heavy-ion collisions and for probing the effects of cold nuclear matter.
 
This chapter provides a pedagogical introduction of the theory and phenomenology of fragmentation functions. We begin in Sec. 2 with the formal field-theoretical definition of FFs, discussing their key properties such as sum rules and their scale dependence as governed by the DGLAP evolution equations. In Sec. 3, we review the primary experimental processes used to probe FFs, including electron-positron annihilation, semi-inclusive deep inelastic scattering, and proton-proton collisions. Sec. 4 summarizes the methodology of global QCD analyses, which combine world data to extract the universal FFs, and compares results from modern analysis groups. The discussion then broadens in Secs. 5 and 6 to extended topics, including dihadron FFs and the three-dimensional picture of hadronization provided by spin- and transverse-momentum-dependent FFs. Finally, Sec. 7 explores the modification of fragmentation within a nuclear medium before we conclude with a summary and outlook on future directions in the field.

\section{Definition and properties of unpolarized fragmentation functions}\label{sec:def-unpol}

\subsection{Field-theoretical definition of FFs}
The operator definition of FFs in QCD provides the field-theoretical foundation for describing hadronization, it is an essential ingredient for the development of QCD factorization framework. The field-theoretical definition of FFs is formalized by Collins and Soper in 1981 \cite{Collins:1981uw}. By defining FFs as specific, gauge-invariant matrix elements of QCD operators, they established FFs as fundamental, universal quantities, which allows the application of perturbative QCD to a vast range of high energy experimental observables for hadron production. To handle the relativistic kinematics of high-energy processes, it is convenient to use light-cone coordinates, where a vector $v^\mu$ is given by $(v^+, v^-, \mathbf{v}_T)$ with $v^{\pm} = (v^0\pm v^3)/\sqrt{2}$. 

The starting point for the definition is the fragmentation correlator $\Delta_i^h(z)$, with $i=u,\bar u,d,\bar d,s,\bar s,c,\bar c \cdots$ stands for the fragmenting partons and $h=\pi^0,\pi^{\pm},K^{\pm},p,\bar p,\eta\cdots$ are the observed hadron, $z$ is the momentum fraction of the fragmenting parton carried by the observed hadron. Specifically, the fragmentation of a quark $q$ with momentum $k$ into a hadron $h$ with momentum $P_h$ and spin $S_h$ is described by a bilocal fragmentation correlator. This correlator represents the interference of amplitudes where a parton fragments into a specific hadron plus unobserved remainders $X$ \cite{Collins:1981uw,Collins:2011zzd,Mulders:1995dh,Bacchetta:2006tn}
\begin{equation}
\Delta_q^h(z, P_h, S_h) = \sum_X \int \frac{d^3\vec{P}_X}{(2\pi)^3 2P_X^0} \int \frac{dy^-}{2\pi} e^{ik^+ y^-}
\langle 0 | \mathcal{W}(\infty^-, y^-) \psi(y^-) | P_h, S_h; X \rangle
\langle P_h, S_h; X | \bar{\psi}(0) \mathcal{W}(0, \infty^-) | 0 \rangle_{y^+=0, \mathbf{y}_T=0},
\end{equation}
where the Wilson line is in the fundamental representation, expressed as a path-ordered exponential of the gluon field, $\mathcal{W}(y_1, y_2) = \mathcal{P} \exp\left[-ig \int_{y_1}^{y_2} dy'_\mu A^\mu(y')\right]$. It connects the two field operators, acting as a ``color transporter'' to ensure that the quark fragmentation correlator is physically meaningful and independent of the choice of gauge. The sum over all out states $X$ other than the observed $h$ in the operator definition is equivalent to ensuring we are considering the inclusive number of hadrons produced, regardless of what else is created in the fragmentation process.

The correlator $\Delta_q^h$ is a $4 \times 4$ matrix in Dirac space and contains information about the spin configurations of the fragmenting quark and the observed hadron. In general, this matrix can be decomposed into a basis of 16 independent Dirac structures. At leading power in the energy scale, three of them are multiplied by the corresponding scalar fragmentation functions at leading twist. The general decomposition reads
\begin{equation}
\Delta_q^h(z, P_h, S_h) = \frac{1}{z} \left[ D_q^{h}(z)\,\gamma^- - \lambda_h G_{q}^{h}(z)\,\gamma^-\gamma_5 + S_{hT,i} H_q^{h}(z)\,i\sigma^{i-}\gamma_5 + \cdots \right].
\end{equation}
Here, $D_q^{h}(z)$ is the unpolarized FF, $G_{q}^{h}$ is the helicity FF probed by the hadron's longitudinal polarization $\lambda_h$, and $H_q^{h}$ is the transversity FF probed by the hadron's transverse polarization $S_{hT}$. The ellipsis denotes other fragmentation structures at higher twist. While the polarized FFs are essential for understanding spin physics and will be introduced in Sec. \ref{sec-spin}, the unpolarized FF, $D_i^h$, is the most well studied quantity describing the hadron production rates and will be the focus of this chapter.

To isolate the unpolarized collinear fragmentation function, we average over the spin of the initial parton and sum over the spin of the final observed hadron. The term ``collinear" signifies that we integrate over the transverse momentum of the hadron relative to the fragmenting parton, effectively treating it as moving parallel to the parent parton. This approximation simplifies the FF into a 1-dimensional function depending solely on the longitudinal momentum fraction $z$. The unpolarized quark FF $D_q^{h}(z)$ is obtained by taking the trace of the general correlator with the Dirac matrix $\gamma^+$. This projection isolates the desired spin-averaged probability density, leading to the full operator definition of quark FFs \cite{Collins:1981uw,Collins:2011zzd,Mulders:1995dh,Bacchetta:2006tn}
\begin{equation}
D_q^{h}(z) = \frac{z}{4} \sum_{X, S_h} \int \frac{d^3\vec{P}_X}{(2\pi)^3 2P_X^0} \int \frac{dy^-}{2\pi} e^{i k^+ y^-} \text{Tr} \left[ \gamma^+ \langle 0 | \mathcal{W}(\infty^-,y^-) \psi(y^-) | P_h, S_h; X \rangle \langle P_h, S_h; X | \bar{\psi}(0) \mathcal{W}(0,\infty^-) | 0 \rangle \right]_{y^+=0, \mathbf{y}_T=0}.
\label{eq-Dq}
\end{equation}
This expression formally defines the number density of unpolarized hadrons of type $h$ emerging from the fragmentation of an unpolarized quark of flavor $q$, carrying a fraction $z$ of the quark's momentum. The dependence on the factorization scale $\mu^2$ is implicit and arises from the renormalization of this operator as shown in Sec. 2.3.

The definition of the gluon FF, $D_g^{h}(z)$, follows the same logic but uses the gluon field strength tensor $F^{\mu\nu}$ and a Wilson line in the adjoint representation, reflecting the gluon's color-octet nature. The operator definition for gluon FF reads \cite{Collins:1981uw,Collins:2011zzd,Mulders:1995dh,Bacchetta:2006tn}
\begin{equation}
D_g^{h}(z) = \frac{z}{2k^+} \sum_{X, S_h} \int \frac{d^3\vec{P}_X}{(2\pi)^3 2P_X^0}  \int \frac{dy^-}{2\pi} e^{i k^+ y^-} \langle 0 | F^{+i}(y^-) \mathcal{W}_{\rm adj}(\infty^-,y^-) | P_h, S_h; X \rangle \langle P_h, S_h; X | \mathcal{W}_{\rm adj}(0,\infty^-) F^{+i}(0) | 0 \rangle_{y^+=0, \mathbf{y}_T=0}.
\label{eq-Dg}
\end{equation}

The profound importance of these operator definitions lies in their role in proving QCD factorization theorems. Because the FFs are defined as matrix elements describing the transition from vacuum to hadronic state initiated by a fundamental field operator, their properties are independent of how that parton was created. This means, for example, the function $D_u^{\pi^+}(z)$ describing an up-quark fragmenting into a positive pion is the exact same universal function whether the up-quark was produced in electron-positron annihilation, deep-inelastic lepton-proton scattering, or a proton-proton collision.
This essentially means that $D_i^h$ can be measured in one process and used as an input to make predictions for another. This enables quantitative comparisons between QCD theory and a vast array of experimental data.

\subsection{Interpretation and properties of FFs}

While the operator definition of FFs provides a rigorous field-theoretical foundation, the physical meaning of fragmentation functions is best understood through their connections to probabilities, particle numbers, and the fundamental conservation laws.

The physical interpretation of FFs can be understood using the language of light-front quantization \cite{Collins:1981uw,Collins:2011zzd}, where the physical out states can be built by creation operators acting on a vacuum. The number operator for a specific hadron $h$ is constructed from its creation ($a_h^\dagger$) and annihilation ($a_h$) operators:
\begin{equation}
    \hat{N}_h = \int \frac{dP_h^+ d^2\mathbf{P}_{h\perp}}{(2\pi)^3 2P_h^+} a_h^\dagger(P_h) a_h(P_h)
    =\sum_{X,S_h}\int \frac{dP_h^+ d^2\mathbf{P}_{h\perp}}{(2\pi)^3 2P_h^+} | P_h, S_h; X \rangle \langle P_h, S_h; X |.
\end{equation}
Implement the above equation to the operator definitions of FFs as shown in Eqs. (\ref{eq-Dq}) and (\ref{eq-Dg}), one can understand the fragmentation function as the differential expectation value of this number operator in a state created by a parton field operator acting on the vacuum, $| \text{parton state} \rangle \equiv \bar{\psi}(0) |0\rangle$. Schematically, the number of hadrons with momentum $P_h$ is given by the expectation value $\langle \text{parton state} | a_h^\dagger(P_h) a_h(P_h) | \text{parton state} \rangle$. Therefore, the FF $D_q^{h}(z)$ is precisely the distribution of this hadron number as a function of the momentum fraction $z = P_h^+/k^+$, i.e. $d\langle \hat{N}_h \rangle /dz \propto D_q^{h}(z)$, which results in the total hadron number sum rule of FFs:
\begin{equation}
    \langle N_h \rangle = \int_0^1 dz D_i^h(z). 
\end{equation}
This connects the abstract operator definition to the concrete, physical picture of counting particles produced from a partonic source. 

On the other hand, intuitively from the process of a fragmentation, the total momentum of the initial fragmenting parton must be carried by the ensemble of all produced hadrons. In terms of the creation and annihilation operators, the momentum operator can be written as 
\begin{equation}
    \mathcal{P}^{\mu} = \sum_h\int \frac{dP_h^+ d^2\mathbf{P}_{h\perp}}{(2\pi)^3 2P_h^+} a_h^\dagger(P_h) P_h^{\mu} a_h(P_h).
\end{equation}
Substitute this definition into the matrix element $\langle \text{parton state} | a_h^\dagger(P_h) \mathcal{P}_h^+ a_h(P_h) | \text{parton state} \rangle$, which can be written as the sum of the momentum of each hadron type, weighted by its multiplicity, leading to the momentum sum rule:
\begin{equation}
    \sum_h \int_0^1 dz \, z D_i^{h}(z, \mu^2) = 1.
\end{equation}
Again, the sum runs over all possible hadron species $h$ of the second moment, and it is valid for any flavor $i$. One can prove that this momentum sum rule is preserved after renormalization, especially works in the commonly used renormalization schemes like modified minimal subtraction scheme (${\rm \overline{MS}}$). 

The conservation of the electromagnetic current requires that the net electric charge must be conserved during hadronization. The total charge of the produced hadrons must equal the charge of the fragmenting parton, $Q_i$. The derivation can be performed in a similar fashion as in the case for momentum conservation, one just need to replace the momentum operator by the charge operator, leading to the charge sum rule:
\begin{equation}
    \sum_h Q_h \int_0^1 dz\, D_i^{h}(z, \mu^2) = Q_i,
\end{equation}
where $Q_h$ is the charge of the detected hadron $h$. For a gluon ($Q_g=0$), this implies that the net charge of the hadrons from gluon fragmentation must be zero. Similarly, using isospin SU(2) symmetry for the favored and unfavored fragmentation functions, and also the charge conjugation invariance of the strong interactions, one can derive the generalized conservation law for any other conserved quantum number, such as the strangeness and baryon number.  

Establishing the theoretical foundations of the sum rules involves navigating a subtle conceptual ``paradox" regarding how a colored initial state evolves into a colorless final state, and the associated formal derivation remains an open question. Intuitively, the total momentum of an initial fragmenting parton must be carried by the ensemble of all produced hadrons. However, a significant conceptual hurdle exists: a fragmenting quark carries color and fractional quantum numbers (like electric charge), whereas all observed final-state hadrons are color-singlet objects with integer charges. One possible explanation for how these mismatched states are equivalent is provided by Collins and Rogers \cite{Collins:2023cuo}. They point out that a gauge-invariant definition of a fragmentation function requires a Wilson line, which acts as an external source of color charge. In a confining theory, this implies the final state must include not only ``normal" hadrons but also an ``orphan" state—a bound state between the Wilson line and a soft parton from the vacuum that neutralizes the system's color. While the ``orphan" state is necessary to conserve internal symmetries like color and charge, it is kinematically distinct from the observed hadrons. Because the Wilson line extends to infinity, these anomalous bound states are restricted to a fractional momentum of essentially zero ($z=0$). Consequently, the contribution from the ``orphan" state vanishes in the momentum integral. This explains why the momentum sum rule remains valid for observed hadrons, even if other sum rules, such as those for electric charge or baryon number, require further investigations.

The sum rules of FFs are expected to provide important consistency checks and constraints for phenomenological models and global fits of fragmentation functions. However, the practical application of these sum rules, primarily the momentum sum rule as shown in Eq. (8), into the realistic global analyses is a challenge. This is mainly caused by the fact that the summation over all hadron species is not accessible in existing experiments, and the integration over all momentum faction regions down to $z=0$ can not be reliably obtained, as well as the non-convergence of perturbative QCD series in low-$z$ region. Nevertheless, taking the sum rules as a priori boundary constraints, several groups have performed phenomenological extractions of FFs, such as \cite{deFlorian:2017lwf,deFlorian:2014xna,Albino:2008fy,Hirai:2007cx,Kretzer:2000yf}. And global analyses of FFs have also been obtained without such a priori constraint but rather is used as a posteriori check \cite{Bertone:2017tyb,Gao:2024nkz,Gao:2025bko}.

\subsection{QCD evolution of unpolarized fragmentation functions}

As shown in the operator definitions of FFs, the involved hadronic scale indicates the non-perturbative nature of FFs. In the naive parton model, FFs are scale-invariant (a property known as scaling), depending only on the momentum fraction $z$. However, in QCD, an energetic parton can radiate gluons or split into quark-antiquark pairs before it hadronizes. This quantum-mechanical process introduces a logarithmic dependence on the energy scale $\mu$ (also denoted as factorization scale $\mu_f$ in the framework of QCD factorization), a phenomenon known as scaling violation. While the $z$ dependence of FFs is not predictable, their dependence on energy scales can be determined by perturbative QCD (pQCD), provided $\mu\gg \Lambda_{\rm QCD}$.

\begin{figure}[h]
	\centering
	\includegraphics[width=1\textwidth]{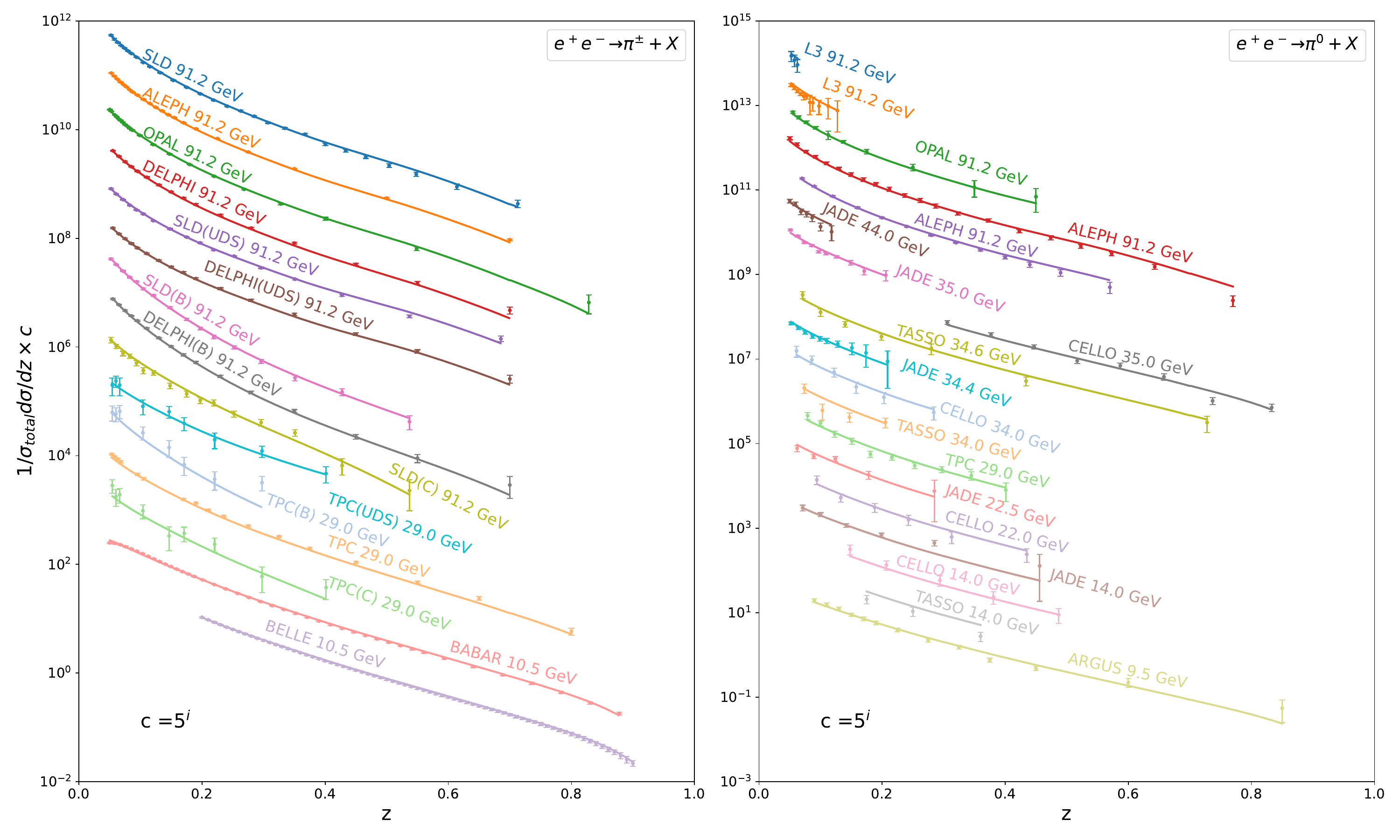}
	\caption{Comparison of world data sets for pion production and theoretical calculations at NNLO accuracy in various energies. The distributions have been scaled, with $i$ ranging from 0 for Belle to 14 for SLD for $\pi^{\pm}$ (left) production \cite{SLD:1998coh,ALEPH:1994cbg,OPAL:1994zan,DELPHI:1998cgx,TPCTwoGamma:1988yjh,Belle:2013lfg,BaBar:2013yrg}, and from 0 for ARGUS to 18 for L3 for $\pi^{0}$ (right) production \cite{L3:1991nwl,L3:1994gkb,OPAL:1998enc,ALEPH:1996oqp,ALEPH:1999udi,JADE:1989ewf,CELLO:1989byk,TASSO:1986tou,JADE:1985bzp,TASSO:1981rff,CELLO:1983mif,TPCTwoGamma:1984eyl,ARGUS:1989orf}. The figure is taking from Ref. \cite{Li:2024etc}.}
	\label{fig:SIA-data}
\end{figure}

This scale dependence is governed by the famous DGLAP evolution equations~\cite{Gribov:1972ri,Gribov:1972rt,Lipatov:1974qm,Dokshitzer:1977sg,Altarelli:1977zs}, a powerful tool to connect measurements performed at different energy scales.
The DGLAP equations describe how the probability of a parton fragmenting into a given hadron changes as the energy scale $\mu$ of the probes. For unpolarized, leading-twist FFs $D_i^h(z, \mu^2)$, the evolution equation is an integro-differential equation:
\begin{equation}
\frac{\partial D_i^h(z, \mu^2)}{\partial \ln \mu^2} = \sum_j \int_z^1 \frac{dy}{y} P_{ji}\left(y, \alpha_s(\mu^2)\right) D_j^h\left(\frac{z}{y}, \mu^2\right),
\label{eq-DGLAP}
\end{equation}
where the sum runs over all parton types $j$ (quarks, antiquarks, and gluons) in the intermediate splitting process. The physical interpretation of DGLAP evolution shown above can be easily read that the parton $j$ with momentum fraction $y$ fragments into hadron $h$, with $j$ itself came from a parent parton $i$ via a splitting process, as encoded by the splitting function $P_{ji}(y, \alpha_s)$. This time-like splitting function represents the probability density for parton $i$ to split into parton $j$. The crucial insight of the DGLAP framework is that the splitting functions $P_{ji}$ are universal and can be calculated systematically in pQCD as a power series in the strong coupling constant $\alpha_s$:
\begin{equation}
P_{ji}(y, \alpha_s) = \sum_{n=0}^\infty \left( \frac{\alpha_s}{2\pi} \right)^{n+1} P_{ji}^{(n)}(y).
\end{equation}
These splitting functions have been computed to high precision, see e.g. the leading order (LO) results are obtained in Refs. \cite{Gribov:1972ri,Gribov:1972rt,Lipatov:1974qm,Dokshitzer:1977sg,Altarelli:1977zs}, the next-to-leading order (NLO) coefficients have been calculated in Refs.~\cite{Curci:1980uw,Furmanski:1980cm,Floratos:1981hs,Kalinowski:1980ju}, and the next-to-next-to-leading order (NNLO) corrections ~\cite{Mitov:2006ic,Moch:2007tx,Almasy:2011eq} represent the highest precision in current stage. 

While the DGLAP equations for timelike evolution of FFs and spacelike evolution of PDFs are basically identical, their evolution kernels are different. According the Gribov-Lipatov relation \cite{Gribov:1972rt}, the LO splitting functions for timelike and spacelike evolutions are equivalent, i.e. $P_{ji}^{(0)}\leftrightarrow P_{ij}^{(0)}$. However, this relation does not hold beyond LO, instead, their connections can be formulated by analytic continuation \cite{Chen:2020uvt}.

The DGLAP evolution predicts that as the energy scale $\mu$ increases, partons have more ``phase space" for radiation. This leads to a characteristic pattern of scaling violations, and the probability of producing a hadron with a large momentum fraction $z$ decreases, while the probability of producing softer hadrons at small $z$ increases. This happens because a high-energy parton is more likely to undergo multiple splittings ($q \to qg$, $g \to gg$, $g \to q\bar{q}$), degrading its energy before the final hadronization step. This precise pattern has been observed by experiments. For instance, in $e^+e^-$ annihilation as shown in Fig. \ref{fig:SIA-data} for pion production, the measured distributions become noticeably softer as the center-of-mass energy increases from the scales of Belle to those of SLD measurements. These measurements provide one of the most compelling quantitative tests of pQCD.

The DGLAP evolution of fragmentation functions plays a significant role in QCD phenomenology. It provides the essential theoretical machinery to relate non-perturbative hadronization, measured at a certain energy scale, to predictions at any other high-energy scales. By parametrizing FFs at a single initial scale $\mu_0$ shown in Sec. 4, their form at any scale $\mu > \mu_0$ is completely determined by QCD. This predictive power as shown by the solid curves in Fig. \ref{fig:SIA-data}, and its successful verification by experiment, validates the factorization framework and allows for precision studies of hadron production across all high-energy scales.

\section{Experimental probes of fragmentation functions}
\begin{figure}[h]
	\centering
	\includegraphics[width=0.9\textwidth]{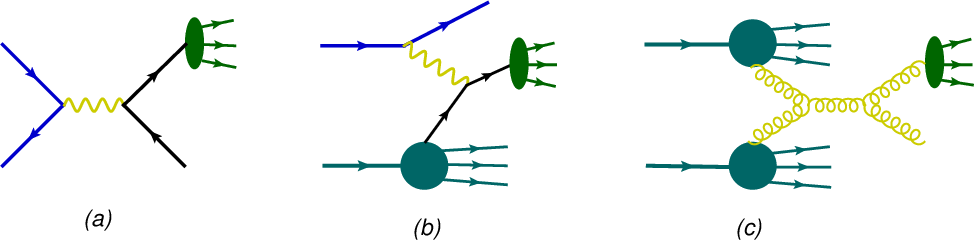}
	\caption{Schematic diagrams for processes involving FFs: (a) electron-positron annihilation,  (b) semi-inclusive deep inelastic scattering of lepton-hadron collision, and (c) proton-proton collision.}
	\label{fig:Feyn-diam}
\end{figure}

Since fragmentation functions are non-perturbative and cannot yet be calculated directly from first principles, they must be extracted from experimental data. The universality of FFs, a critical property in the framework of QCD factorization for inclusive hadron production, allows for a global analysis that combines world data from several different types of high-energy scattering processes. The three primary processes for constraining FFs are shown in Fig. \ref{fig:Feyn-diam}, including single inclusive hadron production in electron-positron annihilation (SIA), semi-inclusive deep inelastic scattering (SIDIS) of lepton-hadron collisions, and proton-proton collisions. Each process offers unique sensitivities to different aspects of hadronization, such as parton flavor and spin dependence. 
\subsection{Hadron production in electron-positron annihilation}
Among the three primary processes for probing FFs, SIA process as shown in Fig. \ref{fig:Feyn-diam}(a), $e^+ + e^- \rightarrow \gamma^*/Z \rightarrow h + X$, provides the cleanest and most direct probe of fragmentation functions. The main advantage of this probe arises from its purely leptonic initial state, which is precisely understood within quantum electrodynamics. SIA process avoids the complications associated with the nonperturbative internal structure of initial-state hadrons, such as PDFs and beam remnants involved in lepton-hadron and hadron-hadron collisions. This theoretical clarity makes SIA an ideal laboratory for performing high-precision extractions of FFs \cite{Gao:2025hlm,Bertone:2017tyb,Soleymaninia:2018uiv,Borsa:2022vvp,AbdulKhalek:2022laj} and testing the underlying theorem of QCD factorization \cite{BESIII:2024hcs,Li:2024etc}.

The SIA cross section for producing a single inclusive hadron $h$ at large momentum transfer $Q$ (set by the center-of-mass energy $\sqrt{s}$) can be expressed in terms of fragmentation structure functions $F_{T,L,A}^h$ \cite{Nason:1993xx,Webber:1994zd}
\begin{equation}
\frac{1}{\sigma_0}\frac{d\sigma^h}{dxd{\rm cos}\theta} = \frac{3}{8}(1+{\rm cos}^2\theta)F_T^h(x,Q^2) + \frac{3}{4}{\rm sin}^2\theta F_L^h(x,Q^2) + \frac{3}{4}{\rm cos}\theta F_A^h(x,Q^2),
\end{equation}
where $\theta$ denotes the angle of hadron with respect to the electron beam direction in the center of mass frame, $x=2E_h/\sqrt{s}$ with $E_h$ the energy of the detected hadron. The structure functions $F_T^h$ and $F_L^h$ represent the contributions from transverse and longitudinal polarizations of the exchanged $\gamma$ or $Z$-boson, while the asymmetric structure function $F_A^h$ arises from the interference between vector and axial-vector contributions, which vanishes in the integration over all $\theta$ angles. 

According to the leading twist QCD factorization, the fragmentation structure functions $F_{T,L,A}^h$ can be systematically separated into a short-distance partonic cross section $C_i$ and the long-distance FFs:
\begin{equation}
F_{T,L,A}^h = \sigma_0\sum_i \int_x^1 \frac{dz}{z} \, C^i_{T,L,A}(z,\alpha_s,Q^2,\mu^2) \, D_i^h\left(\frac{x}{z},\mu^2\right),
\end{equation}
where $\sigma_0$ is the Born-level cross section for $e^+e^- \to \mu^+\mu^-$. The coefficient functions $C^i$ describe the hard scattering subprocess and can be calculated in pQCD depending on the factorization scheme, which is typically chosen to be the $\overline{\text{MS}}$ scheme. Remarkable precision of these perturbative calculations have been carried out at NLO~\cite{Altarelli:1979kv,Baier:1979sp,deFlorian:1997zj}, NNLO~\cite{Rijken:1996vr,Rijken:1996npa,Rijken:1996ns,Mitov:2006ic,Mitov:2006wy}, and even next-to-next-to-next-to-leading order (N$^3$LO)~\cite{He:2025hin}. These high precision calculations significantly reduce the theoretical uncertainty and provides essential theoretical input for global extraction of FFs from world data.

Despite its cleanliness, inclusive SIA has limitations in its sensitivity to the flavor structure of FFs. Because the intermediate virtual photon or Z boson couples to quarks and antiquarks in an approximately symmetric way, these measurements are primarily sensitive to the flavor-singlet combination of FFs, namely $D_q^h(z,\mu^2) + D_{\bar{q}}^h(z,\mu^2)$. This makes it difficult to constrain charge asymmetries or to separate the contributions of individual quark and antiquark flavors. Furthermore, since gluons are not produced at leading order in $e^+e^-$ annihilation, sensitivity to the gluon FF $D_g^h$ arises only from higher-order corrections, as encoded in the scaling violation of the DGLAP equations, or the specific multi-jet event topologies.

The experimental foundation for our knowledge of FFs has been built over several decades by a succession of $e^+e^-$ colliders. In particular, the legacy data from the experiments at LEP (ALEPH, DELPHI, L3, OPAL) and the Standford Linear Collider (SLD) at the $Z^0$-pole provided high-precision measurements of various charged hadron spectra. The $B$-factories, Belle at KEK and BaBar at SLAC, operating at a center-of-mass energy of $\sqrt{s} \approx 10.58$ GeV, produced enormous datasets that allowed for unprecedented precision in FF measurements. The excellent particle identification capabilities in these experiments enabled separate determination of FFs for pions, kaons, and protons over a large momentum range. These facilities also pioneered the first measurements sensitive to spin and transverse-momentum-dependent FFs, such as through Collins and dihadron asymmetries as will be introduced later. To overcome the limitations of traditional SIA measurements and enhance the sensitivity to flavor decomposition of FFs, experimental analyses have employed more sophisticated techniques. The use of flavor-tagged cross sections allows for the statistical separation of events originating from light quarks ($u,d,s$), charm quarks ($c$), and bottom quarks ($b$). This is achieved using event-shape variables or, more powerfully, through the identification of secondary vertices from the long lifetimes of heavy flavor hadrons. Additionally, the study of three-jet events, particularly those where one jet can be identified as a gluon jet through its distinct properties, provides a more direct access to gluon fragmentation. The longitudinal fragmentation structure function, $F_L$, which only arises at order $\mathcal{O}(\alpha_s)$ and beyond, is also sensitive to the gluon FF, though it is relatively more difficult to measure than that for the transverse structure function.

More recently, the BESIII experiment, operating at lower energies in the $\tau$-charm threshold region ($\sqrt{s} = 2.0-4.9$ GeV), has provided valuable new information. These measurements are used to test the validity of QCD factorization in low energy regime, where effects such as hadron masses and higher twist corrections can become important \cite{Li:2024etc}. The inclusion of BESIII data in global fits helps to better constrain the low-scale behavior of FFs and has revealed some tension with extrapolations from high-energy data \cite{Gao:2025hlm}, highlighting necessity for theoretical improvement~\cite{He:2025hin}.

Historically, many other experimental measurements of SIA also provided significant input for the constrain of FFs. More complete collections of data sets can be found in Ref. \cite{Metz:2016swz} and Durham HEP data base \cite{HepData}. In addition, the proposed future facilities, such as ILC \cite{Bambade:2019fyw}, CEPC \cite{CEPCStudyGroup:2018rmc,CEPCStudyGroup:2018ghi}, FCC-$ee$ \cite{FCC:2018evy}, and STCF~\cite{Achasov:2023gey} will provide more opportunities for the probing of FFs in a vast range of energies from a few GeV to TeV scales.

\subsection{Hadron production in semi-inclusive deep inelastic scattering}
SIDIS process as shown in Fig. \ref{fig:Feyn-diam}(b), $\ell(l) + N(P) \rightarrow \ell'(l') + h(P_h) + X$, serves as an indispensable complement to SIA in the global study of fragmentation. By using nucleon or nucleus as a target, SIDIS introduces a dependence on the PDFs of initial nucleon or nucleus, which, while adding complexity, provides the crucial leverage needed for a unique flavor decomposition of the FFs.

For an unpolarized lepton beam scattering off an unpolarized nucleon target, following the notation in Ref. \cite{Anderle:2012rq} and considering only virtual photon exchange, the cross section of SIDIS can be written as:
\begin{equation}
\frac{d\sigma^{\ell N \to \ell' h X}}{dx_B \, dy \, dz_h} = \frac{4\pi\alpha^2}{Q^2} \left[\frac{1 + (1 - y)^2}{2y} \mathcal{F}_T^h(x_B,z_h,Q^2) +\frac{1-y}{y}\mathcal{F}_L^h(x_B,z_h,Q^2)\right],
\end{equation}
where $\alpha$ denotes the fine structure constant, $x_B=Q^2/(2P\cdot q)$ is the Bjorken scaling variable probing the momentum fraction of the struck quark inside the nucleon, and $z_h=P\cdot P_h/(P\cdot q)$ is the momentum fraction of the struck quark carried by the final-state hadron. $Q^2=-q^2$ with $q$ denotes the momentum of the exchanged virtual photon, and $y=P\cdot q/(P\cdot \ell)$ represents the so called inelasticity. 

In the framework of leading twist QCD collinear factorization, the transverse $\mathcal{F}_T^h$ and longitudinal $\mathcal{F}_L^h$ structure functions can be factorized into the convolution of perturbative coefficient functions $C_k^{ij}$ and nonperturbative PDFs $f_{i/N}$ and FFs $D_j^h$. 
\begin{equation}
 \mathcal{F}_k^h(x_B,z_h,Q^2) = \sum_{i,j}\int_{x_B}^1 \frac{dx}{x}\int_{z_h}^1\frac{dz}{z} f_{i/N}(x_B/x,\mu_f^2)C^{ij}_k(x,z,Q^2,\mu_f^2)D_j^h(z_h/z,\mu_f^2),  ~~~~k=T,L
 \label{eq-SIDIS}
\end{equation}
While the coefficient $C^{ij}_k$ have been carried out at NLO~\cite{Baier:1979sp,Altarelli:1979kv,deFlorian:1997zj} more than four decades ago, the NNLO calculations were realized only recently~\cite{Goyal:2023zdi,Bonino:2024qbh}. Such NNLO results were implemented for global extraction of charged hadron FFs at NNLO accuracy for the first time in Ref. \cite{Gao:2025hlm}.

The key advantage of SIDIS process lies in the strong correlation of PDF $f_{i/N}$ with parton flavor $i$ and FF $D_j^{h}$ with flavor $j$, as shown in the above factorization formula Eq. (\ref{eq-SIDIS}). Though this inherent entanglement with PDFs challenges analyzing SIDIS data, which introduces an additional layer of non-perturbative uncertainty into the extraction of FFs. Fortunately, the PDFs for initial nucleon are much better constrained comparing to FFs of any hadron. Therefore, the nucleon target can act as a ``flavor filter" for the flavor decomposition of FFs. By measuring the production rates of different identified hadrons (e.g., $\pi^{\pm}$, $K^{\pm}$, $p/\bar p$) from both proton and deuteron targets, global analyses can statistically disentangle the fragmentation functions for up, down, and strange quarks and antiquarks. This is particularly vital for separating the unfavored fragmentation processes (e.g. $d \to \pi^+$) from the favored ones (e.g. $u \to \pi^+$) and for determining the fragmentation functions of strange hadrons. Favored fragmentation occurs when the starting parton is a valence quark of the final-state hadron (e.g. $u \to \pi^+$  or $d \to \pi^-$). These processes generally have higher probabilities because the parton can directly transition into the hadron's primary structure. Unfavored fragmentation describes cases where the parton is not a valence constituent (e.g. $d \to \pi^+$ or $s \to \pi^+$ ), requiring the creation of the necessary quark-antiquark pairs from the vacuum (sea). On the other hand, SIDIS process plays a similar role with SIA in probing gluon FFs, since it only enters at NLO and not at LO, making SIDIS process insensitive to the constrain of gluon fragmentation. Furthermore, SIDIS is the primary experimental tool for accessing the rich spin and transverse-momentum structure of hadronization. By polarizing the lepton beam and/or the nucleon target, one can measure a wide variety of spin asymmetries that provide access to polarized and TMD FFs. 

A series of fixed-target experiments have provided the crucial data for SIDIS-based fragmentation studies. The EMC experiment \cite{EuropeanMuon:1991sne} at CERN and HERMES experiment \cite{HERMES:2012uyd} at DESY, with its variety of proton, hydrogen, deuterium, and nuclear targets, provided pioneering measurements of hadron multiplicities that established the methodology for flavor decomposition. The COMPASS experiment \cite{COMPASS:2016xvm} at CERN, using a high-energy muon beam, extended these measurements to smaller values of $x_B$ and larger values of $Q^2$, and was particularly influential in its measurements of polarized asymmetries. While with higher energy lepton-proton colliders at HERA \cite{ZEUS:1995acw,H1:1996muf}, substantially increased multiplicities have been measured and provided complementary observation of scaling violation of FFs. More recently, the 12 GeV upgrade at Jefferson Lab has enabled a program of high-precision SIDIS measurements with the Hall A/B/C spectrometers. These experiments provide unprecedented data in the high-$x$ valence quark region, offering precise constraints on the large-$z$ behavior of FFs and testing the limits of the factorization framework.

The future of the field lies with the planned Electron-Ion Collider in US (EIC) \cite{AbdulKhalek:2021gbh} and the proposed Polarized Electron-Ion Collider in China (EicC) \cite{Anderle:2021wcy}. As the next generation high-luminosity electron-ion machines, the EIC and EicC will represent new opportunities in our ability to study hadronization. Its wide kinematic coverage and high precision will allow for a definitive mapping of the flavor dependence of FFs, including those for sea quarks and gluons, with unparalleled accuracy.

\subsection{Hadron production in proton-proton collisions}
Hadron production in proton-proton ($pp$) collisions as shown in Fig. \ref{fig:Feyn-diam}(c), the third pillar of fragmentation studies, offers unique and powerful access to certain partonic channels, most notably gluon fragmentation. While the theoretical description is significantly more complex than for leptonic collisions due to the presence of two initial-state hadrons, the high energy coverage of colliders like the Relativistic Heavy Ion Collider (RHIC) and the Large Hadron Collider (LHC) makes these processes essential for a complete understanding of FFs.

A key advantage of $pp$ collisions is that gluon-initiated subprocesses (e.g., $gg \to gg$, $gq \to gq$) contribute at the leading order of perturbation theory and are often the dominant production channels at high energies. This is in contrast to SIA and SIDIS, where gluon FFs appear only at higher orders or through scaling violations. Consequently, measurements at hadron colliders provide the most direct and sensitive constraints on the gluon fragmentation function $D_g^{h}$.

The most traditional observable in $pp$ collisions is single inclusive hadron production, $p + p \rightarrow h + X$, at high transverse momentum $p_{T}^h$. The factorized cross section involves a double convolution over the PDFs of both initial protons, followed by a convolution with the FFs:
\begin{equation}
E_h \frac{d\sigma^h}{d^3P_h} = \sum_{a,b,c} \int dx_a \, dx_b \, dz_c \, f_{a/p}(x_a,\mu^2) \, f_{b/p}(x_b,\mu^2) \, \frac{d\hat{\sigma}_{ab\to cX}}{d\hat{t}} \, D_c^h(z_c,\mu^2),
\end{equation}
where $f_{i/p}$ are proton PDFs, and $d\hat{\sigma}$ are the partonic hard-scattering cross sections, which have been calculated analytically at NLO \cite{Aversa:1988vb,Jager:2002xm} and numerically at NNLO \cite{Czakon:2025yti}. 

In recent few years, a more refined and powerful observable has become central to FF analyses, i.e. the measurement of hadrons inside a fully reconstructed jet. This jet fragmentation function is defined as the momentum distribution of hadrons within a jet, typically as a function of $z_h = p_T^h / p_T^{\text{jet}}$. This observable has the advantage of reducing the complexity of the convolution, as the jet's momentum provides a better proxy for the fragmenting parton's momentum and flavor than can be inferred in fully inclusive hadron production. This allows for a more direct and local probing of the FF's dependence on the momentum fraction. Furthermore, advanced techniques are being developed to distinguish quark-initiated jets from gluon-initiated jets, which will provide even cleaner access to FFs with specific flavors. NLO pQCD calculations for jet fragmentation function in $pp$ collisions have been carried in Refs.  \cite{Arleo:2013tya,Caletti:2024xaw,Kaufmann:2019ksh,Liu:2023fsq}, providing essential input for the global analysis of FFs.

The initial, crucial constraints on the gluon FF came from the extensive measurements of single inclusive hadron spectra by the PHENIX and STAR experiments at RHIC (see e.g. Refs. \cite{STAR:2011iap,PHENIX:2011rvu}). These data, at a center-of-mass energy of 200 GeV, provided strong constrain for the shape of $D_g^{h}$. Following this, the experiments at the LHC by ATLAS, CMS, LHCb and ALICE collaborations have provided a wealth of data on inclusive hadron production at much higher energies (see e.g. Refs. \cite{ALICE:2014juv,ALICE:2019hno,ALICE:2020jsh}), up to 13 TeV. The vast kinematic reach of these measurements offers essential input for testing the DGLAP evolution of FFs and has significantly improved their precision.

The most recent and impactful progress, however, has come from the high precision measurements of jet fragmentation functions at the LHC (see e.g. Refs. \cite{CMS:2018mqn,LHCb:2022rky,ATLAS:2011myc}). These measurements combined with pQCD calculations have been implemented in a recent global analysis, which has dramatically improved the precision of all FFs, and particularly that of the gluon~\cite{Gao:2024nkz,Gao:2024dbv}. This highlights the power of jet-based observables in disentangling fragmentation dynamics. The ongoing analysis of these datasets, along with future techniques to statistically distinguish quark- and gluon-initiated jets, continues to be a central focus of the field.

In conclusion, while hadronic collisions present significant theoretical challenges due to their intricate convolution structure and additional non-perturbative effects like the underlying event, the data they provide are indispensable. The unique leading-order access to gluon FFs makes $pp$ collisions a critical and complementary component in all modern global QCD analyses of FFs.

\section{Global analysis of unpolarized fragmentation functions}\label{sec:global-unpol}

The ultimate goal of fragmentation function analysis is to determine a single, universal set of FFs that can consistently describe world data from all relevant high-energy processes. This is achieved through a global QCD analysis, a sophisticated procedure that combines a vast array of experimental data from SIA, SIDIS, and $pp$ collisions within the theoretical framework of QCD factorization and DGLAP evolution.

\subsection{Phenomenological models for hadronization}
\begin{figure}[h]
	\centering
	\includegraphics[width=0.7\textwidth]{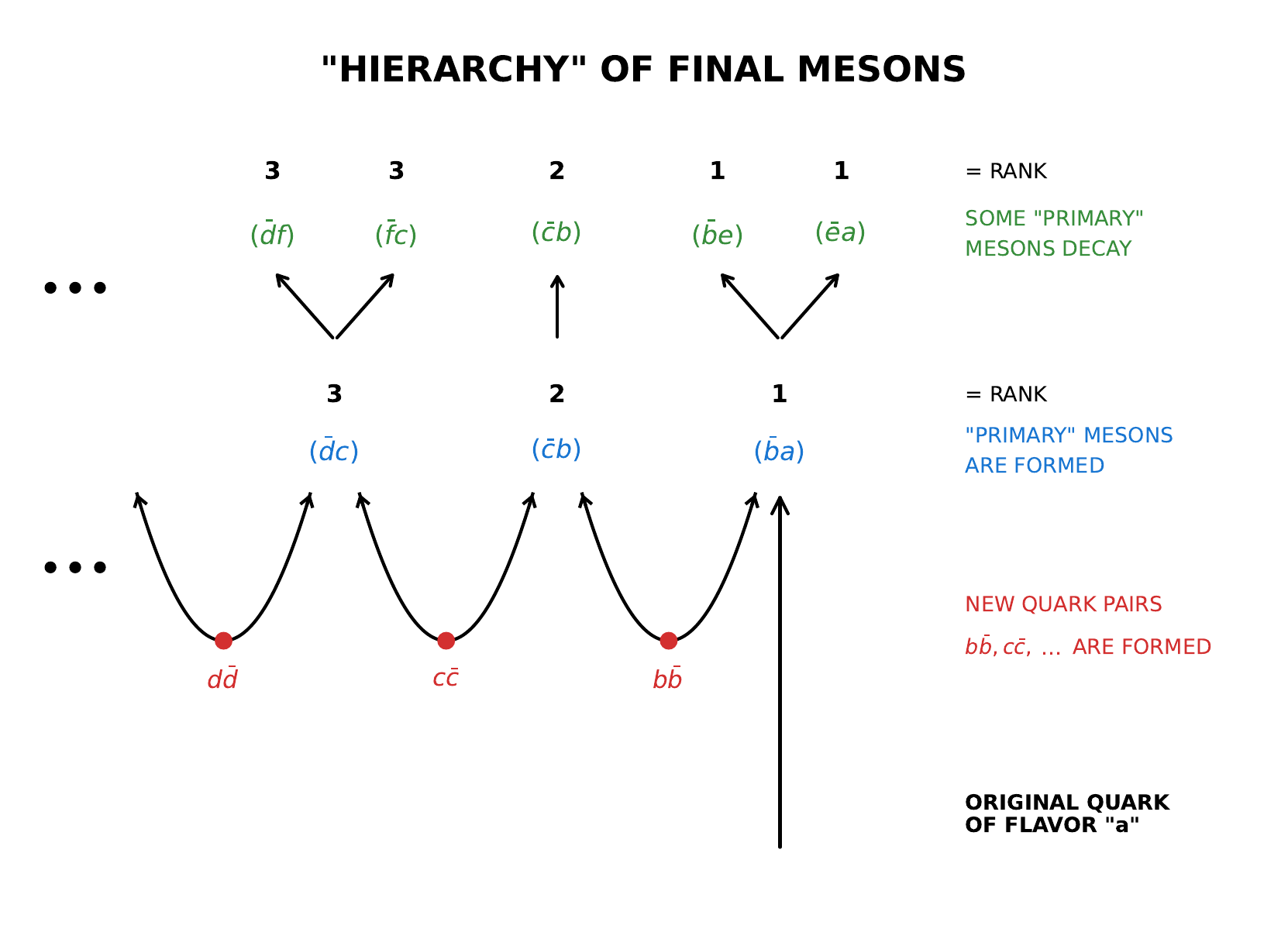}
    	\caption{The recursive ``hierarchy" of meson production in the Feynman-Field fragmentation model. An initial quark of flavor $a$ combines with an antiquark $\bar b$ from a vacuum-produced $q \bar q$ pair to form the ``rank 1" primary meson ($a \bar b$). The leftover quark $b$ repeats the process, creating a ``rank 2" meson ($b \bar c$), and so on, until the available energy is depleted. Unstable primary mesons subsequently decay into final-state hadrons, which inherit the rank of their parent. (Figure adapted from Refs. \cite{Field:1977fa}).}
	\label{fig:string}
\end{figure}
Before the era of rigorous QCD-based global fits, the understanding of parton fragmentation was guided by intuitive phenomenological models. The most influential of these was the Feynman-Field model~\cite{Field:1977fa}, which pictured fragmentation process as a recursive, scale-invariant cascade. In this model, as shown in Fig. \ref{fig:string}, a quark fragments into a color singlet hadron by combining with another antiquark from a $q\bar q$ pair created from vacuum. The leftover quark then repeat the previous procedure of fragmentation, until the available energy is depleted. Once the primary mesons are formed through the iterative $q\bar q$ pair creation, those that are unstable undergo further decays based on established branching ratios. The process of quark fragmentation was described by a simple parametrized function, $f(\eta)$, representing the probability that the leftover quark retains a momentum fraction $\eta$ of its parent. The fragmentation function was then constructed as
\begin{equation}
    D(z) = f(1-z) + \int_z^1 \frac{d\eta}{\eta} f(\eta) D\left(\frac{z}{\eta}\right).
\end{equation}
The first term represents the probability that the meson is created in the first step fragmentation, while the second term represents the recursive probability for the successive fragmentations. The parameters in $f(\eta)$ are determined by comparing to experimental data, similar to the global determination of FFs at the initial scale. Although superseded by the DGLAP formalism, the Feynman-Field model introduced the core concepts of universality and scaling, and its recursive structure laid the conceptual groundwork for modeling the hadronization processes in modern Monte Carlo event generators.

The most evident example is the hadronization models used by event generators like PYTHIA \cite{Bierlich:2022pfr}, Herwig \cite{Bahr:2008pv} and Sherpa \cite{Gleisberg:2008ta}. These programs first use pQCD to generate a parton shower, describing the evolution from a high energy scale down to a lower non-perturbative scale. Then, the showered partons hadronize using specific models. The Lund String Model \cite{Andersson:1983ia}, used in PYTHIA, models the color field between a separating quark-antiquark pair as a one-dimensional string. As the partons move apart, the potential energy stored in the string increases until it breaks the string  by producing a new $q\bar{q}$ pair from the vacuum, creating colorless hadrons. The Cluster Model \cite{Webber:1983if}, used in Herwig and Sherpa, first forces all gluons to split into $q\bar{q}$ pairs, then groups color-singlet quark-antiquark pairs into ``clusters" which then decay into the observed hadrons according to phase space. While more sophisticated, both models retain the core idea of a local, iterative process that converts partonic energy into hadrons, as inherited from the Feynman-Field picture.

\subsection{Methodology of global analysis}
Global analysis of FFs is a sophisticated methodology to extract  fundamental information about parton hadronization and to interpret experimental results at high energies. It is a complex task involving experimental measurements,  QCD theoretical calculations and computing science, aiming to find a set of non-perturbative functions that best describes all relevant data. In general, a global analysis includes the following key steps:
\begin{enumerate}
    \item \textbf{Data collection:} A primary goal of FF global extractions is the determination of fragmentation functions for each individual quark flavors ($u$, $d$, $s$, $c$, $b$), antiquarks, and gluons. Such a flavor separation is a challenge because most observables are sensitive to a sum over all flavors. Disentangling these contributions requires combining the unique sensitivities of different processes in the data collection process. As clarified in previous section, electron-positron annihilation is primarily sensitive to the flavor-singlet sum of quark and antiquark FFs with charge-squared weighting. Measurements that tag the flavor of the initiating quark (e.g., heavy-flavor tagging in SLD \cite{SLD:2003ogn} and DELPHI \cite{DELPHI:1998cgx} experiment) provide crucial separated information. Semi-inclusive deep inelastic scattering provides an access for flavor separation. By using different targets (proton vs. deuteron) and detecting different final-state hadron species, one can take advantage of the known flavor structure of the proton's PDFs to isolate the FFs of individual light quark flavors. Proton-proton collisions provide the strongest constraints on the gluon FF, which is dominant at high energies. So far, a wide-range of dataset from SIA, SIDIS, and $pp$ experiments, covering scattering energies from a few GeV to TeV, are assembled. Stringent kinematic cuts should be applied to each dataset to ensure the measurements are in a kinematic region where QCD factorization is expected to hold and fixed order QCD corrections are convergent, e.g. sufficiently high $Q\gg \Lambda_{\rm QCD}$ and $z\ge 0.05$.

    \item \textbf{FF parametrization:} The exact functional form for FFs is unknown due to their non-perturbative nature. Therefore, a parametrization form is presumed to mimic the unknown FFs at the initial scale $\mu_0$. This is a crucial step where different strategies exist and theoretical bias can be introduced. The traditional approach is to use a flexible polynomial form, typically parametrized as follows
    \begin{equation}
        D_i^{h}(z, \mu_0^2) = N_i^h z^{\alpha_i^h} (1-z)^{\beta_i^h} f_i^h(z).
    \end{equation}
    The parameters $\{N_i^h, \alpha_i^h, \beta_i^h\}$ are free parameters of the fit, to be determined for each parton and hadron species. The above functional form is inspired by theoretical expectations for the behavior of FFs at the endpoints $z \to 0$ and $z \to 1$. The $z^\alpha$ term models the evolution behavior of small-$z$, while the $(1-z)^\beta$ term controls the large-$z$ fall-off, motivated by dimensional-counting rules \cite{Jones:1978he}. The function $f_i^h(z)$ was set to enhance the flexibility of parametrizing FFs at initial scale, which was set to 1 in early fits as motivated by the Feynman-Field model \cite{Field:1977fa} when the data constraining power was limited, see e.g. \cite{Ali:1979em}. The extension of $f_i(z)$ to more complicated polynomial functional forms were widely used and have been effectively model the FFs for various species of hadrons. This setup may not capture all the features presented in the data, potentially underestimating the true uncertainty of the FFs. To mitigate the bias introduced by a fixed functional form, the degree of polynomials in $f_i^h(z)$ has been increased till no significant improvements of fit are observed, with the final values varying from 0 to 2 depending on the flavors of parton and hadron \cite{Gao:2024nkz}. 

    Alternatively, to provide a sufficient region of function space and reduce parametrization bias, machine learning techniques, particularly neural networks, have been increasingly adopted. This approach was first established for PDFs by the NNPDF collaboration and has been successfully extended to FFs~\cite{Bertone:2017tyb}. In this framework, the fragmentation function for each parton is parametrized by a multi-layer feed-forward neural network. The momentum fraction $z$ serves as the input, and the output, after processing, gives the value of $zD(z, \mu_0^2)$. The network consists of interconnected nodes (``neurons'') arranged in layers, with the connections having associated ``weights'' and each neuron having an ``activation function''. These weights are the parameters of the fit. With a sufficient number of nodes, a neural network can act as a universal approximator, capable of learning any smooth function from the data without a pre-conceived shape.

    Unlike parton distribution functions, which are limited to the proton, the fragmentation functions involve various species of hadrons, leading to a large sets of FFs (e.g., even with just pions, there exists more than 20 FFs). Therefore, it is usually necessary to apply some symmetries among a large sets of FFs, such as the isospin SU(2) symmetry for the favored and unfavored fragmentation functions, and charge conjugation invariance of the strong interactions (e.g., $D_q^{\pi^+,K^+,p} = D_{\bar q}^{\pi^-,K^-,\bar p}$). These presumed symmetries can significantly reduce the total amount of independent fragmentation functions. 
    
    \item \textbf{QCD evolution and prediction:} For each data point measured at scale $\mu$, the parametrized FFs are evolved from an initial scale $\mu_0$ to the experimental scale $\mu$ using the DGLAP equations. These are a set of coupled integro-differential equations, shown in Eq. (\ref{eq-DGLAP}), that describe how the fragmentation functions change with the energy scale. The solution of these equations can be performed using different methodologies, primarily in Mellin space or in $z$-space (also called $x$-space).
    
    In the DGLAP equation, the involved convolution integral form over the momentum fraction $z$ can be solved directly in $z$-space. This involves discretizing the momentum fraction $z$ on a grid and solving the integro-differential equations numerically. The convolution integrals are then replaced by sums over the discretized grid points. Numerical methods like Runge-Kutta are commonly employed to integrate the evolution equations with respect to $\ln Q^2$. The $z$-space approach is generally more flexible regarding the functional form of the parametrized FFs, as it doesn't require knowing their Mellin transforms. However, it can be computationally more intensive due to the explicit calculation of convolutions. The Mellin transform, i.e. $\tilde D(N) = \int_0^1 dz z^{N-1} D(z) $, offers a powerful simplification by converting convolution integrals into simple products. The DGLAP equations in Mellin space are generally easier to solve analytically or numerically. Once the Mellin moments are evolved to the desired $Q^2$ scale, an inverse Mellin transform is performed to return the FFs to $z$-space. This approach is often numerically efficient, especially for higher-order QCD calculations, as it avoids complex numerical convolutions.

    Several public programs are available to solve the DGLAP equations for both PDFs and FFs, implementing either Mellin space or $z$-space methodologies. APFEL (A PDF Evolution Library with QED corrections) \cite{Bertone:2013vaa} is a widely used library that solves DGLAP equations in $z$-space with high precision, including QCD corrections up to NNLO and QED corrections. It is written in Fortran 77 using higher-order interpolation and Runge-Kutta solvers, and can also be used via the C/C++ and Python interfaces. An accompanied package called MELA (Mellin Evolution LibrAry) is designed in Mellin-space, aiming to provide a simple and user-friendly cross-check of the $z$-space evolution in APFEL. Another Fortran 77 program called QCDNUM program \cite{Botje:2010ay} also solves the evolution equations directly in $z$-space for FFs in perturbative QCD up to NNLO, where a evolution algorithm based on quadratic spline interpolation yields good performance in accuracy and speed. A versatile convolution engine is also provided in this program to compute cross-sections in various scattering processes. HOPPET (Higher Order Perturbative Parton Evolution Toolkit) \cite{Salam:2008qg} is a commonly used Fortran 95 package that performs DGLAP evolution in $z$-space using Runge-Kutta techniques, providing tools for manipulating PDFs and FFs on a grid with good speed and accuracy. QCD-PEGASUS \cite{Vogt:2004ns} is a well-known program that performs DGLAP evolution in Mellin space, then inverts back to $z$-space. It provides fast and accurate solutions of the evolution equations up to NNLO QCD corrections, but the initial distributions are needed in a form facilitating the computation of the complex Mellin moments.
        
    Based on the DGLAP evolution and the factorization formalisms shown in previous section for various observables, the theoretical predictions can be performed by convoluting the evolved FFs with the appropriate partonic cross sections and, in the case of SIDIS and $pp$, with PDFs. 
    
    \item \textbf{$\chi^2$ minimization and uncertainty estimation:} A global goodness-of-fit function, typically a $\chi^2$, is constructed to quantify the agreement between theory and data, properly accounting for statistical and all correlated and uncorrelated systematic uncertainties. The free parameters of the FF parametrization are then varied to find the minimum of this $\chi^2$, performed using the minimization programs like MINUIT \cite{James:1975dr}. Once the best-fit parameters are found, the uncertainty on the resulting FFs can be estimated. Reliable estimation of uncertainties is an important aspect of any global QCD analysis, these uncertainties arise from multiple sources, including both statistical and systematic experimental errors, theoretical uncertainties, such as those from missing higher-order corrections, estimated by scale variations. Two main frameworks are used to quantify these uncertainties, including Hessian method \cite{Pumplin:2000vx,Pumplin:2001ct} and Monte Carlo sampling methods \cite{Giele:2001mr,Forte:2002fg}.

The Hessian method explores the parameter space around the best-fit minimum of the global $\chi^2$ function. Assuming the $\chi^2$ function is approximately quadratic around its minimum, the curvature of the $\chi^2$ surface at its minimum is described by the Hessian matrix $H_{ij} = \partial^2\chi^2/\partial a_i \partial a_j$, where $a_i$ are the best-fit parameters. A steeper curvature in a particular direction indicates that the parameters associated with that direction are more precisely determined by the data. The inverse of the Hessian matrix provides information about the variances of individual parameters and the correlations between pairs of parameters. A confidence interval is defined by allowing the $\chi^2$ to increase by a specific tolerance, $\Delta\chi^2$, from its minimum value. For example, in a one-parameter fit, a $\Delta\chi^2 = 1$ typically corresponds to a $68\%$ confidence level. This defines a region in the parameter space where the fits are considered statistically acceptable, and the variation of observables within this region constitutes their estimated uncertainty. To find the FF uncertainty for any physical observable $\mathcal{O}$, like a cross-section, one needs to first calculate it with the central FF ($\mathcal{O}({D_0})$) and with each error FF set ($\mathcal{O}(D_k^{\pm})$), where $D_k^{\pm}$ are the FF sets displaced along the $k$-th eigenvector direction of Hessian matrix $H_{ij}$. The total uncertainty is then found by combining the individual variations in quadrature:
    \begin{equation}
    (\Delta\mathcal{O})^2 = \sum_{k=1}^n\left(\mathcal{O}(D_k^+)-\mathcal{O}(D_k^-)\right)^2.
    \end{equation}
    This provides a robust and efficient way to determine the impact of FF uncertainties on theoretical predictions for single inclusive hadron production,  where a relatively small number of error FF sets ( $n$ is typically a few dozen) that are sufficient to determine the uncertainty for any observable.

    The Monte Carlo replica method involves generating an ensemble of pseudo-data sets (``replicas'') by fluctuating the experimental data points according to their uncertainties. Each replica is then fitted independently, producing a full set of FFs. The final central value and uncertainty for any quantity are then the mean and standard deviation of the results from the ensemble of replicas:
    \begin{equation}
    \langle \mathcal{O} \rangle = \frac{1}{N_{\text{rep}}} \sum_{k=1}^{N_{\text{rep}}} \mathcal{O}(D_k), \quad
    (\Delta \mathcal{O})^2 = \frac{1}{N_{\text{rep}}-1} \sum_{k=1}^{N_{\text{rep}}} \left( \mathcal{O}(D_k) - \langle \mathcal{O} \rangle \right)^2.
    \end{equation}
    The Monte Carlo method naturally captures non-Gaussian behavior and provides a robust estimate of uncertainties, especially in regions with limited data. However, this comes at a higher computational cost than the Hessian method, as it typically requires a large number of replicas ($\sim 100-1000$).

\end{enumerate}

\subsection{Comparison of modern fragmentation function sets}

\begin{table}[h!]
\centering
\caption{Comparison of Fragmentation Function Collaborations}
\renewcommand{\arraystretch}{1.5} 
\begin{tabular}{lcccccc}
\hline
\rowcolor[gray]{0.9} 
 & \textbf{NPC} & \textbf{DSS} & \textbf{NNFF} & \textbf{JAM} &  \textbf{MAPFF} \\
\hline
SIA & \checked & \checked & \checked & \checked &  \checked \\
SIDIS & \checked & \checked & \unchecked & \checked &  \checked \\
pp incl. hadron & \checked & \checked & \checked & \checked &  \unchecked \\
pp hadron in Jet & \checked & \unchecked & \unchecked & \unchecked & \unchecked \\
\hline
stat. treatment & Hessian & Hessian & Monte Carlo & Monte Carlo & Monte Carlo \\
parametrization & polynomial & polynomial & neural network & polynomial &  neural network \\
hadron species & \small$\pi^{\pm}, K^{\pm}, p/\bar{p},$ \small$\eta, \pi^0, K_S, \Lambda$ & \small$\pi^{\pm}, K^{\pm}, p/\bar{p}$ & \small$\pi^{\pm}, K^{\pm}, p/\bar{p}$ & \small$\pi^{\pm}, K^{\pm}$ & \small$\pi^{\pm}, K^{\pm}$ \\
pQCD order & NLO/NNLO & NLO/Approx. NNLO & NLO/NNLO & NLO & NLO/Approx. NNLO \\
\hline

\hline
\end{tabular}
\end{table}

Significant progress has been made for the global fits that have achieved a precise extraction of the FFs for different flavors of quarks and gluon into various species of hadrons, including pions, kaons, protons etc. with uncertainties in some cases below 10$\%$ in the well-measured intermediate-$z$ region. In particular, recent progress has also been made in constraining the gluon FF through the inclusion of jet data from the LHC. Nevertheless, the FFs for strange quarks and for all partons at very large and very small $z$ still carry significant uncertainties. Future progress will depend on high-precision data from facilities like the Belle II, EIC and LHC, which will provide the statistics needed to further refine our knowledge of the flavor structure of hadronization.

The landscape of modern FFs is populated by several collaborating groups, each employing distinct methodologies including various experimental, computational, statistical, and theoretical advances, as listed in Tab. I for some representative global analyses. The NPC collaboration \cite{Gao:2024nkz,Gao:2024dbv,Gao:2025hlm,Gao:2025bko} represents the most comprehensive analysis, the key strength is the inclusion of the widest possible range of observables, including SIA, SIDIS, and $pp$ data for both inclusive hadron production and hadron fragmentation in jet from RHIC and the LHC, which provides the most stringent constraints on the gluon FF. In addition, NPC provides as much as possible the hadron species including $\pi^{\pm}, K^{\pm}, p/\bar{p}, \eta, \pi^0, K_S, \Lambda$, with the precision for $\pi^{\pm}, K^{\pm}$ reached to NNLO and for neutral hadrons to NLO accuracy. The DSS group \cite{deFlorian:2007aj,deFlorian:2007ekg,deFlorian:2014xna,deFlorian:2017lwf} and its successors (e.g., BDSS21~\cite{Borsa:2021ran}) represent the traditional approach, using polynomial parametrizations and the Hessian method for uncertainties. The NNPDF collaboration (NNFF for FFs)~\cite{Bertone:2017tyb, Bertone:2018ecm} pioneered the use of neural networks and Monte Carlo replicas for FFs. Their analyses, often based on a more restricted dataset (primarily SIA), aim to minimize parametrization bias and often results in larger, more conservative uncertainties, especially for the less-constrained gluon FF. The JAM collaboration \cite{Sato:2016wqj,Moffat:2021dji} focuses on simultaneous fits of PDFs and FFs, providing a unified picture of both nucleon structure and hadronization. The MAP collaboration \cite{Khalek:2021gxf,AbdulKhalek:2022laj} also employs neural networks but uses the analytic derivative of the network in the fitting process, combining the flexibility of NNs with the efficiency of gradient-based minimization. 

\begin{figure}[h]
	\centering
	\includegraphics[width=.45\textwidth]{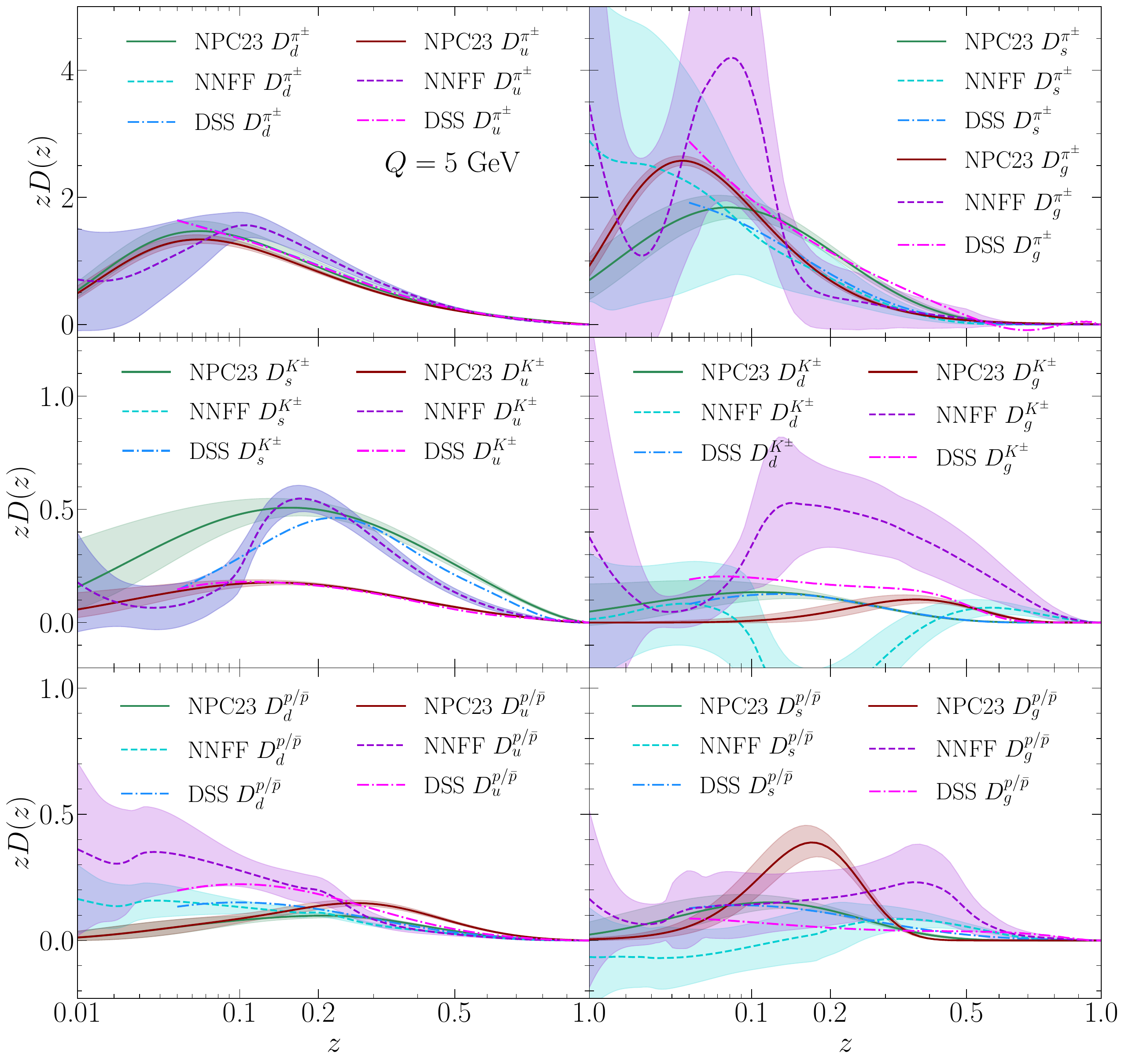}
    \hspace{1cm}
        \includegraphics[width=.44\textwidth]{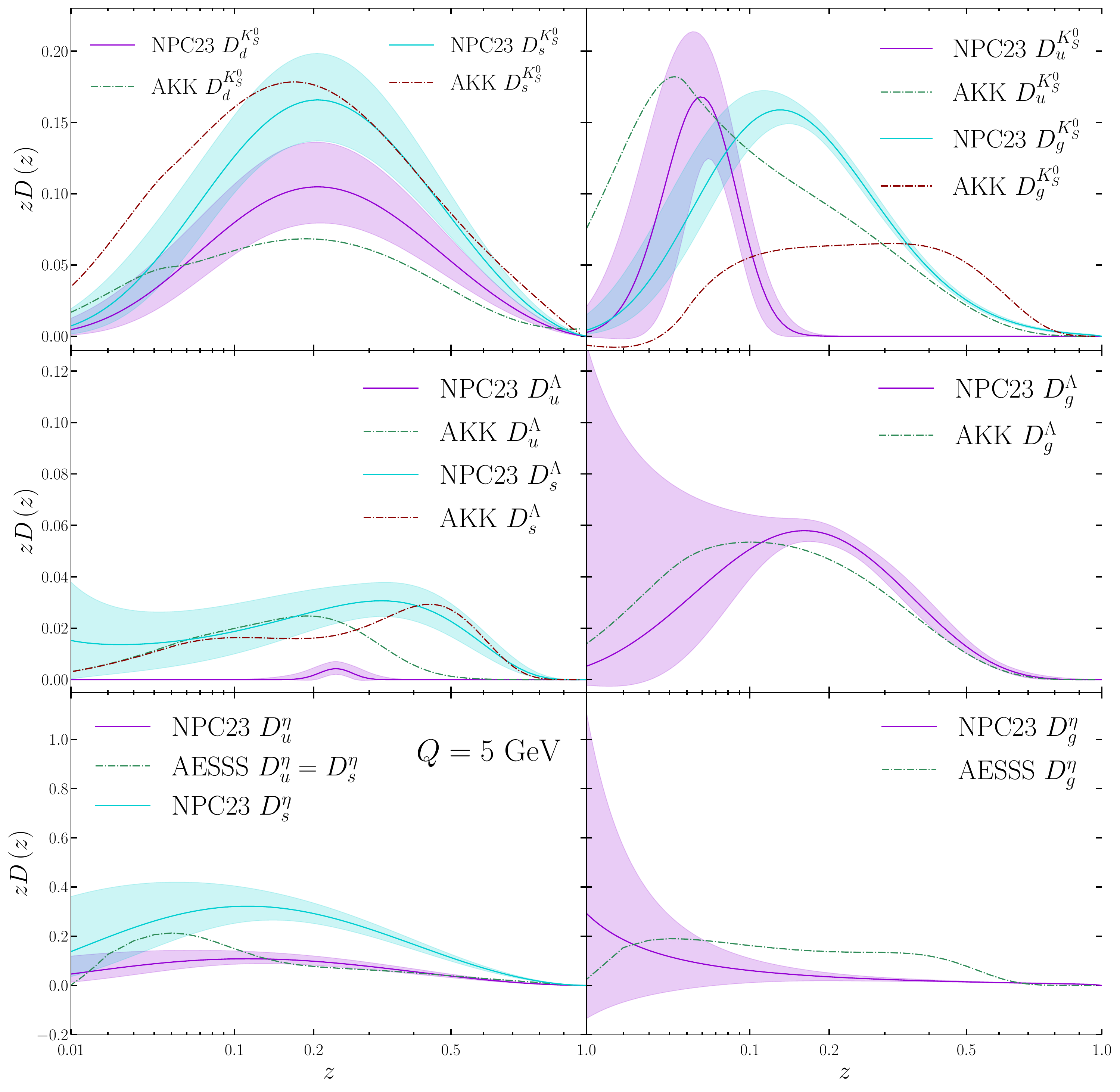}
	\caption{Comparison of fragmentation functions at $Q=5$~GeV. The left panels compare FFs for charged hadrons ($\pi^{\pm},~K^{\pm},~p/\bar p$) from the NNFF, DSS, and NPC23 analyses, uncertainty bands are shown for NNFF and NPC23. The right panels compare FFs for neutral hadrons, showing NPC results against AKK for $K_s^0$ and $\Lambda$, and against AESSS for $\eta$. Figures are adapted from Refs. \cite{Gao:2024nkz,Gao:2025bko}.}
	\label{fig:FF-compare}
\end{figure}

Comparing the results from different groups provides a valuable assessment of the robustness of our understanding on FFs. As an example, we show in the left panels of Fig. \ref{fig:FF-compare} the FFs for charged hadrons $\pi^{\pm}$, $K^{\pm}$, and $p/\bar p$ from NNFF \cite{Bertone:2017tyb}, DSS \cite{deFlorian:2007ekg,Borsa:2021ran,deFlorian:2017lwf}, and NPC23 \cite{Gao:2024nkz,Gao:2024dbv}. This comparison reveals reasonable agreement for the relatively well-constrained $u$ and $d$ quarks to $\pi^{\pm}$. However, significant differences appear for FFs to protons and for FFs of gluon to all three charged hadrons. Notably, the high precision of gluon FFs from NPC global analysis is mostly due to the data of jet fragmentation at the LHC. Furthermore, the newly included data on proton production from SIDIS and pp collisions lead to better flavor separation and thus the differences observed for FFs to protons. Shown in the right panels of Fig. \ref{fig:FF-compare} are the FFs for neutral hadron $K_S^{0}$, $\Lambda$, and $\eta$ from NPC23 \cite{Gao:2025bko}, AKK \cite{Albino:2008fy}, and AESSS \cite{Aidala:2010bn}. For the $d$ and $s$ quark fragmentation to $K_S^{0}$, the results from NPC23 and AKK08 extractions are generally consistent within their respective uncertainties, while the $u$ quark and gluon fragmentation exhibit distinct difference. For $\Lambda$ FFs, the $u$ quark fragmentation from AKK08 are notably larger than NPC23, across the entire $z$ range. As for the fragmentation of gluon and $s$ quark, the results are generally consistent. For the $\eta$ FFs, the shapes and magnitudes of the $u$ quark and gluon fragmentation are generally different between NPC23 and AESSS. The discrepancies among these collaborations highlight the importance of the interplay between different experimental channels and analysis methodologies in achieving a complete and reliable picture of parton fragmentation.

\section{Dihadron fragmentation functions}
Dihadron fragmentation functions (DiFFs) generalize the concept of single-hadron FFs by describing the production of a correlated pair of hadrons from a single parton. They provide an alternative tool to study the dynamics of hadronization beyond the single-hadron inclusive level, and offer a unique pathway to access the transversity parton distribution functions within a collinear framework, rather than the more complicated TMD factorization.

The operator definition of DiFFs is a natural extension of the single-hadron case. One considers the same parton correlator, but now with a two-hadron out state $|P_1, P_2\rangle$ in the final state. The key kinematic variables are the momentum fractions of each hadron in the dihadron pair, $z_{1,2} = P_{1,2}^+/k^+$. Alternatively, one also use $z = z_1+z_2$, $\zeta = z_1-z_2$, and the transverse component $R_T$ of the relative momentum $R_T = (P_{1T}-P_{2T})/2$. The full definition of the quark fragmentation correlator for dihadron can be written as \cite{Collins:1993kq,Bianconi:1999cd,Collins:1994ax,Pitonyak:2023gjx,Majumder:2004wh}
\begin{equation}
    \Delta_q^{h_1 h_2}(z, \vec{k}_T; P_1,P_2) = \sum_X \int \frac{d^3\vec{P}_X}{(2\pi)^3 2P_X^0} \int \frac{d\xi^-d^2\vec{\xi}_T}{(2\pi)^3} e^{i (k^+ \xi^--\vec{k}_T\cdot\vec{\xi}_T)} \langle 0 | \mathcal{W}(\infty,\xi) \psi(0^+,\xi^-,\vec{\xi}_T) | P_1, P_2; X \rangle \langle P_1, P_2; X | \bar{\psi}(0^+,0^-,\vec{0}_T) \mathcal{W}(0,\infty) | 0 \rangle,
\end{equation}
where the transverse momentum $k_T$ of the fragmenting quark is kept for a more general discussion. At leading twist, with the identification of $\vec{R}_T$, the decomposition of this correlator yields four DiFFs \cite{Bianconi:1999cd}, providing richer information than the case for single hadron FFs, where only two TMD FFs exist at leading twist. In the collinear case, where the transverse momentum $k_T$ is integrated over, only two DiFFs survive. For an unpolarized quark fragmenting into an unpolarized hadron pair, the trace of the quark fragmenting correlator with $\gamma^+$ gives the unpolarized DiFF, $D_q^{h_1h_2}(z, \zeta, M_h^2)$. For a transversely polarized quark with spin vector $S_T$, the key new object is the interference fragmentation function (IFF), $H_q^{\sphericalangle h_1h_2}(z, \zeta, M_h^2)$, which is extracted via the trace with $i\sigma^{j+}\gamma_5 \epsilon_{T}^{jk} R_{T,k}$, describing the fragmentation of a transversely polarized quark into two unpolarized hadrons. Notice that $M_h$ is the total invariant mass of the hadron pair, which can be related to $\zeta$ and $R_T^2$. If one further integrate over $R_T^2$, only $D_q^{h_1h_2}(z_1, z_2)$ survives, reading as
\begin{equation}
    D_q^{h_1 h_2}(z_1, z_2) = \frac{z}{4}\sum_X \int \frac{d^3\vec{P}_X}{(2\pi)^3 2P_X^0} \int \frac{d\xi^-}{2\pi} e^{i k^+ \xi^-} {\rm Tr}\left[\gamma^+\langle 0 | \mathcal{W}(\infty^-,\xi^-) \psi(0^+,\xi^-,\vec{0}_T) | P_1, P_2; X \rangle \langle P_1, P_2; X | \bar{\psi}(0^+,0^-,\vec{0}_T) \mathcal{W}(0^-,\infty^-) | 0 \rangle\right].
\end{equation}

It has been pointed out that the normalization factor in the above definition should be modified to ensure the physical interpretation of DiFFs in terms of number density, which leads to the hadron number sum rule and the connection with single hadron fragmentation \cite{Majumder:2004br,Pitonyak:2023gjx}:
\begin{eqnarray}
    \sum_{h_1,h_2}\int_0^1 dz_2 \int_0^{1-z_2}dz_1 D_q^{h_1 h_2}(z_1, z_2) = \langle \mathcal{N}(\mathcal{N}-1)\rangle, \\
    \sum_{h_1}\int_0^{1-z_2}dz_1 z_1 D_q^{h_1 h_2}(z_1, z_2) = (1-z_2)D_q^{h_2}(z_2),
\end{eqnarray}
where $\mathcal{N}$ is the total number of hadrons fragmented from quark $q$. The above sum rules can be also applied to gluon DiFFs and  generalized to $n$-hadron fragmentation functions, detailed discussions can be found in \cite{Majumder:2004br,Pitonyak:2023gjx} and references therein. 

The collinear DiFFs obey their own set of DGLAP evolution equations, which is more complex than for single-hadron FFs because the splitting kernels must account for transitions between one- and two-particle states. In particular, the evolution equation for a DiFF is not closed but mixes with the standard single-hadron FFs, forming a larger coupled system. Schematically, for the unpolarized case \cite{Konishi:1979cb,Sukhatme:1980vs,Majumder:2004wh}:
\begin{equation}
\frac{\partial D_i^{h_1h_2}(z_1,z_2, \mu^2)}{\partial \ln \mu^2} = \sum_j \int_z^1 \frac{dy}{y^2} P_{ji}\left(y, \alpha_s(\mu^2)\right) D_j^{h_1h_2}\left(\frac{z_1}{y},\frac{z_2}{y}, \mu^2\right)+\sum_{jk} \int_{z_1}^{1-z_2} \frac{dy}{y(1-y)} \tilde{P}_{ji}^k\left(y, \alpha_s(\mu^2)\right) D_j^{h_1}\left(\frac{z_1}{y}, \mu^2\right)D_k^{h_2}\left(\frac{z_2}{1-y}, \mu^2\right),
\label{eq-DiFF-evo}
\end{equation}
where the first term on the right hand side is commonly called homogeneous term, represents the standard evolution of the two-hadron state and shares the same evolution kernel with single-hadron FFs. The second term is usually called inhomogeneous term, with a different splitting kernel $\tilde{P}$, describes the creation of the dihadron state from two single fragmentation processes, arising from the real emission in parton splittings. The results for the multi-parton splitting functions can be found in Ref. \cite{Chen:2022muj} up to $\mathcal{O}(\alpha_s^2)$. 

The IFF, $H_q^{\sphericalangle h_1h_2}$, is chiral-odd and arises from the interference between different partial waves of the two-hadron system (e.g., s- and p-wave). This interference makes the IFF sensitive to the transverse spin of the fragmenting quark, providing a crucial tool to access the transversity PDF, $h_1(x)$. The evolution equation for $H_q^{\sphericalangle h_1h_2}$ takes the same form as for $D_q^{h_1h_2}$ shown in Eq. (\ref{eq-DiFF-evo}), with the splitting functions replaced by the corresponding transverse polarized case, where the leading order takes the same form as for transversity distribution \cite{Artru:1989zv}.

The experimental access to $D_i^{h_1h_2}$ is the same as those for single hadron FFs $D_i^{h}$. Instead of measuring one single inclusive hadron as discussed in Sec. III, now measure two hadrons on the same side, i.e. $e^++ e^- \to (h_1,h_2)+X$, $e+p\to e'+(h_1,h_2)+X$ and $p+ p \to (h_1,h_2)+X$. The corresponding factorization theorems for dihadron production are also the same as those for single hadron production, but now one need to consider the production of dihadron state from two single fragmentation processes. For example, the factorization for $e^++ e^- \to (h_1,h_2)+X$ can be schematically written as
\begin{equation}
        \frac{d\sigma^{h_1h_2}}{dz_1dz_2} = \sum_{i,j}\sigma^{ij}\otimes D_i^{h_1}\otimes D_j^{h_2} + \sum_i \sigma^{i}\otimes D_i^{h_1h_2}
    \end{equation}

The primary experimental observables for IFF DiFFs are azimuthal asymmetries. In electron-positron annihilation, $e^+e^- \to (h_1 h_2) + (h_3 h_4) + X$, the correlation between the azimuthal orientations of the two back-to-back hadron pairs leads to a $\cos(\phi_R + \phi_{\bar{R}})$ asymmetry proportional to the product $H_q^{\sphericalangle h_1h_2} H_{\bar q}^{\sphericalangle h_3h_4}$, known as Artru-Collins asymmetry \cite{Artru:1995zu}. In SIDIS off a transversely polarized nucleon with spin $S_T$, the azimuthal asymmetry is accessed with respect to $S_T$ and $R_T$. Due to the chiral-odd nature of $H_q^{\sphericalangle h_1h_2}$, it is coupled with the transversity $h_{1q}(x)$, giving rise to an asymmetry in the cross section 
    \begin{equation}
        d\sigma^{h_1h_2} \propto \left( \dots + S_T \sin(\phi_R + \phi_S) \sum_q e_q^2 h_{1q}(x) H_q^{\sphericalangle h_1h_2}(z, M_h^2) \right),
    \end{equation}
where $\phi_R$ is the azimuthal angle of the relative momentum of the hadron pair and $\phi_S$ is the angle of the nucleon's spin. In $pp$ collisions with one proton transversely polarized, the IFF DiFFs can be accessed in the observable of single spin asymmetry measurement
\begin{equation}
        A_{UT} \propto \frac{\sum_{ijk} h_{1i}\otimes f_{j}\otimes \Delta \hat\sigma^{ij\to kl}\otimes H_k^{\sphericalangle h_1h_2}(z, M_h^2) }{\sum_{ijk} f_{i}\otimes f_{j}\otimes \hat\sigma^{ij\to kl}\otimes D_k^{h_1h_2}(z, M_h^2)}.
\end{equation}

Similar to the probe of single hadron fragmentation, $e^+e^-$ data \cite{Belle:2017rwm,Belle:2011cur} provide access to the singlet combination of DiFFs, $ep$ data \cite{HERMES:2008mcr,COMPASS:2023cgk} achieve better flavor separations, and the high energy $pp$ data \cite{STAR:2015jkc,STAR:2017wsi} is expected to be sensitive to the gluon fragmentation. By considering these three primary processes, global analyses of world data have been performed to extract DiFFs, and simultaneously extract of nucleon transversity distribution has also been achieved ~\cite{Radici:2015mwa,Courtoy:2012ry,Cocuzza:2023oam,Cocuzza:2023vqs}. However, the precision of the global analyses is limited comparing to single hadron fragmentation, and more advanced experimental measurements and theoretical development are required in future.

\section{Spin and transverse momentum dependent parton fragmentation}
\label{sec-spin}
A complete description of hadronization requires a three-dimensional (3D) treatment that captures the correlations among the spins and transverse momenta of the fragmenting partons and the produced hadrons. This is achieved through transverse momentum dependent fragmentation functions (TMD-FFs), which provide a more general picture of the fragmentation process than collinear FFs discussed in previous sections. The simpler, one-dimensional collinear fragmentation functions can be understood as a specific case of the general TMD framework.

TMD-FFs retain the parton's transverse momentum ($\vec{k}_T$) relative to the observed hadron, or in a frame of reference in which the fragmenting parton has no transverse momentum, and the produced hadron has transverse momentum $p_{hT}$. Taking quark TMD-FFs as an example, they are defined through the Fourier transform of a non-local quark correlator that depends on both the longitudinal momentum fraction $z$ and the transverse momentum $\vec{k}_T$ \cite{Collins:1981uw,Collins:2011zzd}:
\begin{equation}
\Delta_q^h(z, \vec{k}_T; \mu, \zeta) = \sum_X \int \frac{d^3\vec{P}_X}{(2\pi)^3 2P_X^0}  \int \frac{dy^- d^2\vec{y}_T}{(2\pi)^3} e^{ik^+ y^- - i \vec{k}_T \cdot \vec{y}_T} \langle 0| \mathcal{W}_1(\infty,y) \psi_q(0^+,y^-, \vec{y}_T) |h, X \rangle \langle h, X| \bar\psi_q(0^+,0^-, \vec{0}_T) \mathcal{W}_2(0,\infty) |0 \rangle.
\end{equation}
Different with the purely collinear case, the TMD fragmentation is not a light-cone correlator, and the quark field is also separated along the transverse direction. $\mathcal{W}_i$ denotes the relevant Wilson lines required for gauge invariance, and $\mu$ and $\zeta$ are renormalization and rapidity scales. This correlator can be decomposed using Dirac matrices to define a complete set of TMD-FFs. At leading twist, there are eight such functions, classified by the polarization states of the fragmenting quark and produced hadron (see e.g. reviews \cite{Metz:2016swz,Boussarie:2023izj}):
\begin{center}
\begin{tabular}{lll}
\toprule
\textbf{Hadron Polarization} & \textbf{Quark Polarization} & \textbf{TMD-FF} \\
\midrule
Unpolarized & Unpolarized & $D_q^h(z, k_T^2)$ \\
Unpolarized & Transverse & $H_{1,q}^{\perp\, h}(z, k_T^2)$ (Collins function) \\
Longitudinal & Longitudinal & $G_{1L,q}^h(z, k_T^2)$ \\
Longitudinal & Transverse & $H_{1L,q}^{\perp\, h}(z, k_T^2)$ \\
Transverse & Unpolarized & $D_{1T,q}^{\perp\, h}(z, k_T^2)$  \\
Transverse & Longitudinal & $G_{1T,q}^h(z, k_T^2)$ \\
Transverse & Transverse & $H_{1,q}^h(z, k_T^2)$, $H_{1T,q}^{\perp\, h}(z, k_T^2)$ \\
\bottomrule
\end{tabular}
\end{center}
Among these, the Collins function $H_1^{\perp\, h}$ is chiral-odd and T-odd \cite{Collins:1992kk}, describing the production of unpolarized hadrons from a transversely polarized quark. This function has attracted most of the attention because of its correlation with transversity PDFs of nucleon in certain scattering processes. The FF $D_{1T}^{\perp\, h}$ is also T-odd and describes the production of transversely polarized hadrons from unpolarized quarks.

The previously discussed one-dimensional collinear FFs can be obtained by integrating the TMD-FFs over the transverse momentum $\vec{k}_T$. This integration effectively traces out the transverse information, leaving only the dependence on the longitudinal momentum fraction $z$,
\begin{equation}
D_i^h(z) = z^2\int d^2 \vec{k}_T\, D_i^h(z, k_T^2),~~~~~~~~
G_i^h(z) = z^2\int d^2 \vec{k}_T\, G_{1L,i}^h(z, k_T^2),~~~~~~~~
H_i^h(z) = z^2\int d^2 \vec{k}_T\, \left[H_{1T,q}^h(z, k_T^2)+\frac{k_T^2}{2M_h^2}H_{1T,q}^{\perp h}(z, k_T^2)\right].
\end{equation}
This procedure allows us to define the collinear polarized FFs, which quantify how the spin of a fragmenting parton is transferred to a hadron in this 1-dimensional picture. It is important to note that the T-odd TMD-FFs, like the Collins FFs $H_{1,i}^{\perp\, h}$, have no direct collinear analogue at leading twist because they must average to zero when integrated over all transverse momentum: $\int d^2 \vec{k}_T H_{1,i}^{\perp\, h}(z, k_T^2) = 0$. Their effects are intrinsically linked to the transverse momentum of the produced hadron.

The three primary leading-twist collinear FFs are the unpolarized FF $D_i^h(z)$, the helicity FF $G_{i}^h(z)$, and the transversity FF $H_i^h(z)$. The physical meaning of the unpolarized FFs $D_i^h(z)$ has been introduced. The other two polarized FFs can be understood as number densities in helicity or transversity space. The helicity FF $G_{i}^h(z)$ (sometimes also denoted as $\Delta D_i^h$) describes the fragmentation of a longitudinally polarized quark into a longitudinally polarized hadron, defined as $G_{i}^h(z) \equiv D_{i^+}^{h^+}-D_{i^+}^{h^-}$ \cite{Burkardt:1993zh,Collins:2011zzd,Mulders:1995dh}. It represents the difference between the number of hadrons produced with the same helicity as the parent quark and those with the opposite helicity. $G_i^h(z)$ directly measures the degree of correlation between the longitudinal spins of the fragmenting quark and the produced hadron. The transversity FF $H_q^{h}(z)$ \cite{Jaffe:1996wp}, similarly describes the fragmentation of a transversely polarized quark into a transversely polarized hadron. It quantifies the correlation between the transverse spin of the quark and the transverse spin of the hadron. A key feature of transversity is its chiral-odd nature, making it significantly harder to access experimentally than its helicity counterpart, because the probe of $H_i^h(z)$ must involve another chiral-odd partner to conserve chirality.

The evolution of these functions with the energy scale $\mu$, and their experimental manifestation, differ significantly between the TMD and collinear frameworks. Collinear FFs, such as $D_i^h(z)$, $G_i^h(z)$, and $H_i^h(z)$, obey the DGLAP evolution equations as shown in Sec. II, while the splitting functions are spin dependent, see e.g. \cite{Stratmann:1996hn}. These are integro-differential equations that describe how the FFs change as the energy scale $\mu^2$ increases. The DGLAP equations resum large logarithms of the form $\ln(\mu^2/\Lambda_{QCD}^2)$. The evolution of $D_q^h(z)$ and $G_q^h(z)$ involves mixing with their gluon counterparts, meaning a quark can radiate a gluon, which then fragments. In contrast, because transversity is chiral-odd and there is no gluon transversity distribution, the evolution of $H_q^h(z)$ does not mix with any gluon FFs, which simplifies its evolution pattern. TMD-FFs follow a much more complex evolution framework, governed by the Collins-Soper-Sterman (CSS) formalism \cite{Collins:1984kg}. This is necessary because TMDs depend on two scales: the hard scale $\mu$ and the rapidity scale $\zeta$ that related to the longitudinal momentum. TMD evolution simultaneously handles changes in both scales. For hard scale evolution, the renormalization group equation differs from DGLAP, it describes how the TMD changes with $\mu$. For rapidity evolution, this is unique to TMDs and is described by the Collins-Soper (CS) equation \cite{Collins:1981uk}. It dictates how the shape of the TMD in transverse momentum changes as the energy of the process is varied. The combined evolution equations resum large logarithms of $\ln(Q^2/k_T^2)$, making them essential for describing processes where $k_T \ll Q$. The core of this formalism is the Collins-Soper kernel, which contains non-perturbative information at low transverse momentum that must be modeled or extracted from data, but becomes perturbative and calculable at high momentum. More detailed discussions on TMD evolution for FFs can be found in Ref. \cite{Boussarie:2023izj}.

Experimentally, different observables are required to isolate each FFs. The collinear helicity FFs $G_i^h$ are primarily accessed by measuring the transfer of longitudinal polarization from a lepton or proton beam to final-state hadrons. The most effective probe is the production of $\Lambda$ hyperons, which act as ``self-analyzing polarimeters" because their parity-violating weak decay, $\Lambda\to \pi p$, reveals their polarization. Experiments like OPAL and ALEPH at LEP \cite{OPAL:1997oem,ALEPH:1996oew}, COMPASS at CERN \cite{COMPASS:2009nhs} and HERMES at HERA \cite{HERMES:2006lro}, and STAR at RHIC \cite{STAR:2018pps,STAR:2023hwu} have provided significant constraints on the transfer of polarization from light quarks to $\Lambda$ hyperons. There are also measurements for transversity FFs $H_i^h$ in $e^+e^-$ and $ep$ collisions \cite{OPAL:1997oem,HERMES:2007fpi,Moretti:2018zjm}, and global extractions have been presented in Ref. \cite{Gamberg:2021iat}. 

Measurements on the TMD-FFs have also been preformed in varisous experiments including $e^+e^-$ annihilation, $ep$, and $pp$ collisions. For the unpolarized TMD-FFs $D_i^{h}(z,k_T)$, they are accessed from the transverse momentum distribution of the produced hadron with respect to the thrust axis in $e^+e^-$ annihilation \cite{Belle:2019ywy}, SIDIS process with $p_{hT}\ll Q$ \cite{HERMES:2012uyd,COMPASS:2013bfs}, and jet fragmentation functions in $pp$ collisions \cite{ALICE:2020pga,LHCb:2022rky} based on theoretical development in Refs. \cite{Kang:2019ahe,Kang:2017glf}.
The T-odd Collins Function ($H_{1,i}^{\perp h}$) is probed through single spin asymmetry in the azimuthal distributions of the produced hadron in the hadronic scattering plane. In SIDIS process with a transversely polarized target, the convolution of the transversity PDF $h_i(x)$ and the Collins function results in a characteristic $\sin(\phi_h + \phi_S)$ modulation in the produced hadrons. This ``Collins asymmetry" was observed , see e.g., by HERMES \cite{HERMES:1999ryv,HERMES:2001hbj}, COMPASS \cite{COMPASS:2005csq}, and CLAS \cite{CLAS:2003qum}. The Collins function is also probed in $e^+e^-$ annihilation experiments (Belle \cite{Belle:2005dmx,Belle:2008fdv}, BaBar \cite{BaBar:2013jdt,BaBar:2015mcn} and BESIII \cite{BESIII:2015fyw}), where the production of two back-to-back hadrons is sensitive to the convolution of two Collins functions ($H_{1,i}^{\perp h} \otimes H_{1,\bar i}^{\perp h}$). This allowed for a clean, independent extraction of the Collins function. Combining the $e^+e^-$ and SIDIS data has led to a reasonably well-constrained Collins function and the extraction of the transversity PDF $h_1$ \cite{Kang:2015msa,Anselmino:2015sxa}. Recently, the Collins functions are also probed in STAR measurements through azimuthal transverse single-spin asymmetries for hadron distribution in jets \cite{STAR:2017akg,STAR:2022hqg}.

The study of TMD FFs is a major frontier in hadron physics. While our knowledge is still far from complete, future experiments for SIDIS process at Jefferson Lab, and particularly the EIC will vastly improve our understanding of the complete 3D spin and momentum structure of QCD hadronization.

\section{Parton fragmentation in nuclear medium}
\label{sec:nuclearFFs}
The study of fragmentation functions in nuclear environments is crucial for understanding the modifications to parton hadronization and the QCD dynamics for multiple scattering caused by the presence of nuclear matter. Unlike vacuum fragmentation, where a high-energy parton evolves in isolation, in nuclear systems such as heavy-ion or lepton-nucleus collisions, the hadronization process could occur within a background of cold nuclear matter or even a deconfined quark-gluon plasma (QGP). These nuclear effects lead to significant modifications of FFs, commonly referred to as ``medium-modified fragmentation functions'', encoding the fundamental information of jet transport in the nuclear medium.

One of the signatures of QGP formation in ultra-relativistic heavy-ion collisions is the suppression of high transverse momentum ($p_T$) hadrons in nucleus-nucleus (A+A) collisions compared to $pp$ collisions, after normalizing by the number of binary nucleon-nucleon collisions. This phenomenon, known as ``jet quenching'', is quantified using the nuclear modification factor $R_{AA}$
\begin{equation}
R_{AA} = \frac{1}{N_{\mathrm{coll}}} \frac{d^2N^{AA}/dp_T dy}{d^2N^{pp}/dp_T dy},
\end{equation}
where $R_{AA} < 1$ at high $p_T$ signals parton energy loss effect in the dense QCD medium. This suppression is interpreted as arising from multiple scatterings and induced gluon radiations of the projectile partons as they traverse the hot dense nuclear medium QGP, which can be effectively incorporated into the modified FFs and can differ substantially from their vacuum counterparts.

Several phenomenological frameworks have been developed to incorporate such medium effects into fragmentation functions. In higher-twist formalism, developed within pQCD, the medium modifications arise from power-suppressed, but medium size enhanced, higher-twist contributions associated with multi-parton correlations in nucleus. These lead to medium-induced corrections to the DGLAP evolution equations and can be resummed into the effective medium-modified FFs \cite{Guo:2000nz,Wang:2001ifa}. Alternatively, using soft-collinear effective theory with Glauber gluons, multiple soft scatterings are incorporated to describe coherent gluon emission and energy loss in dense media, and a generalization of the DGLAP evolution equations for FFs to include final-state medium-induced parton shower corrections for large $Q^2$ processes is also proposed in \cite{Chien:2015vja}. 

A variety of observables are sensitive to medium-modified FFs. For single inclusive hadron spectra, the suppression of hadron yields at high $p_T$ in A+A collisions is a direct manifestation of jet quenching and reflects modifications to FFs due to energy loss. For hadron-in-jet distributions, measurements of the distribution of hadrons within reconstructed jets in heavy-ion collisions can reveal softening and broadening of fragmentation patterns compared to $pp$ collisions. Processes like $Z$+hadron or $\gamma$+hadron production allow relatively clean access to trigger parton fragmentation without the uncertainty from trigger bias, since the $Z$-boson or photon escape from the medium with negligible strong interaction. For dihadron and jet-hadron correlations, azimuthal correlations between pairs of high-$p_T$ hadrons or between jets and hadrons are sensitive to medium-induced broadening and energy redistribution in the fragmentation process. For heavy-flavor fragmentation, charm and bottom quarks offer additional sensitivity to the QCD predicted flavor hierarchy of jet quenching due to their finite mass effect, probed via $D$ meson and $B$ hadron production, see e.g. Ref. \cite{Xing:2019xae}.

In the absence of a hot dense medium, fragmentation functions are assumed to be effectively modified in lepton-nucleus ($eA$) and proton-nucleus ($pA$) collisions for single inclusive hadron production due to cold nuclear matter effects, in analogy to the well known effect for nuclear modified parton distribution functions. This idea was first proposed for collinear FFs to explain the production yields for specific identified hadron in a nuclear environment \cite{Sassot:2009sh}, and further explored in \cite{Zurita:2021kli}. Recently, renewed analysis was performed in \cite{Doradau:2024wli} to include world data on the nuclear modification factors in $eA$ from HERMES \cite{HERMES:2009uge} and CLAS \cite{CLAS:2021jhm} and $pA$ collisions from PHENIX \cite{PHENIX:2006mhb}, STAR \cite{STAR:2003oii}, ALICE \cite{ALICE:2016dei} and LHCb \cite{LHCb:2022tjh}. 
The obtained nuclear modified FFs favor a strong suppression for quark fragmentation and moderate enhancement for gluon fragmentation, and both effects growing with nuclear size $A$.  

\begin{figure}[h]
	\centering
    \includegraphics[width=.4\textwidth]{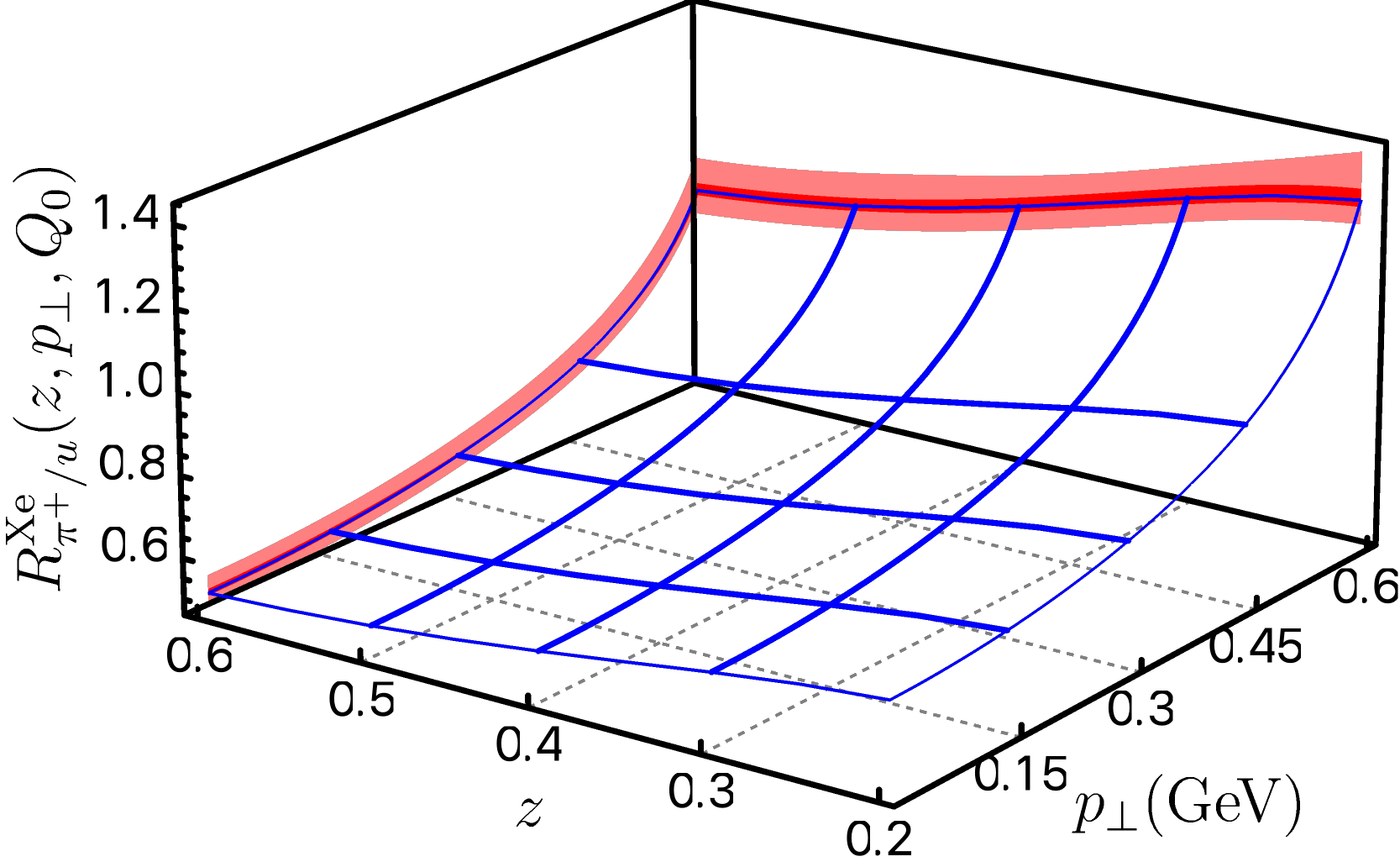}
    \hspace{1.5cm}
    \includegraphics[width=.4\textwidth]{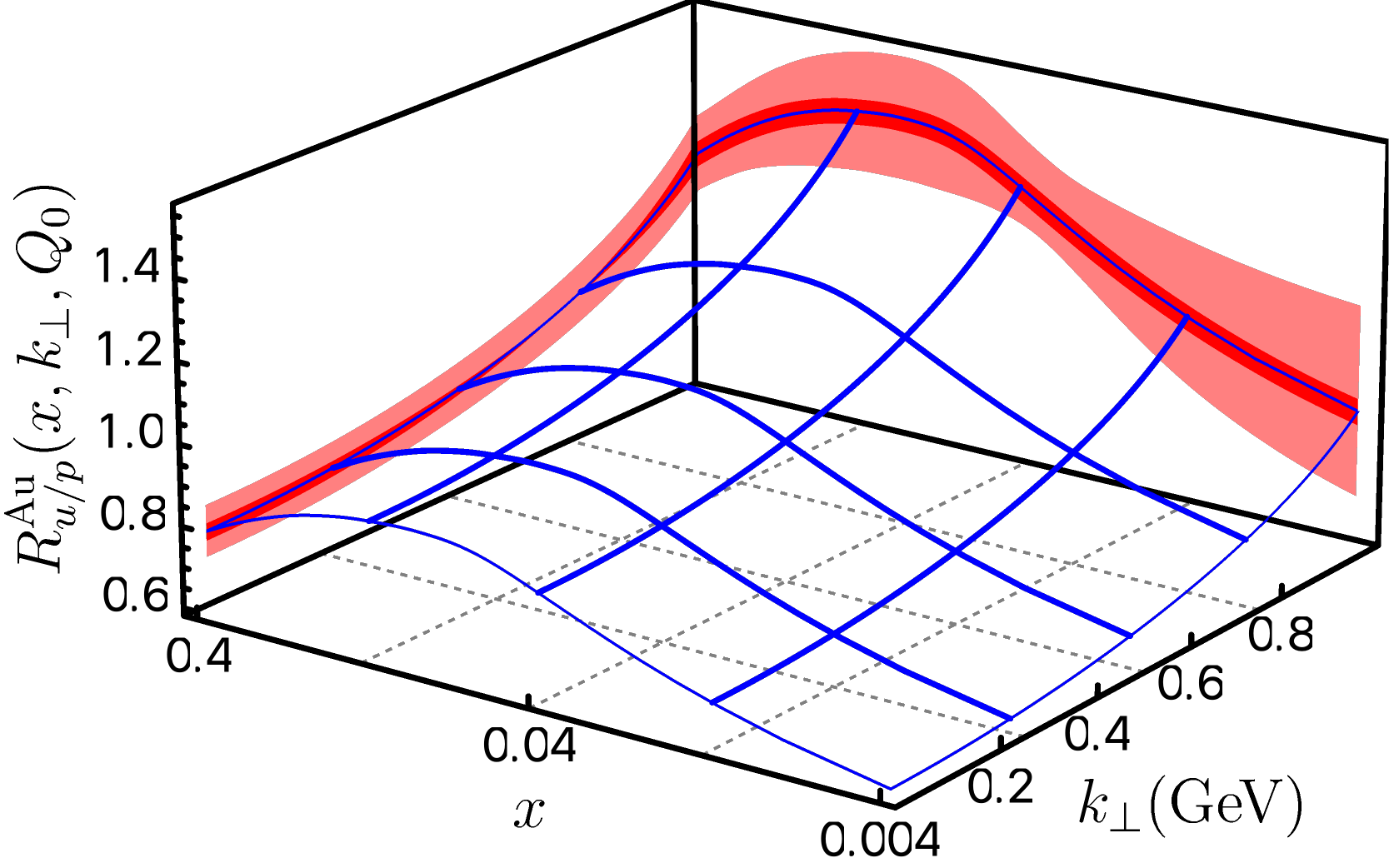}
	\caption{The 3D ratio $R_{h/i}^A(z,p_{\perp},\mu) = D_{i/A}^h(z,p_{\perp},\mu)/D_i^h(z,p_{\perp},\mu)$ for valence quark as a function of $z$ and $p_{\perp}$ (left), the counter part 3D ratio plot is also shown for the nuclear modified TMDPDFs (right) \cite{Alrashed:2021csd} .}
	\label{fig:nTMDFF}
\end{figure}

The idea of nuclear modified FFs was extended, for the first time, to the transverse momentum dependent case in Ref. \cite{Alrashed:2021csd}, i.e. three-dimensional imaging of parton fragmentation in nuclei. A simultaneous global QCD extraction of TMD parton distribution functions and the TMD fragmentation functions (nTMDFF) in nuclei was performed \cite{Alrashed:2021csd,Alrashed:2023xsv}, including world data from SIDIS in HERMES \cite{HERMES:2007plz} and CLAS \cite{CLAS:2021jhm}, and Drell-Yan dilepton production at Fermilab \cite{Alde:1990im,NuSea:1999egr}, RHIC \cite{Leung:2018tql}, and LHC \cite{ATLAS:2015mwq,CMS:2015zlj}. At next-to-leading order and next-to-next-to-leading logarithmic accuracy, as shown in Fig. \ref{fig:nTMDFF} for the ratio of the nTMDFF for $u \to \pi$ in Xe (left plot), one finds the quark fragmentation in medium favors a more broadly distributed feature in transverse momentum than in vacuum, similar with the phenomenon for counterpart nTMDPDFs (right plot).

The study of parton fragmentation in nuclear media remains a key topic in nuclear physics. With ongoing and future experiments at RHIC, the LHC, and the EIC, increasingly precise measurements will further constrain the modification mechanisms of FFs and deepen our understanding of hadronization in QCD matter. 

\section{Conclusions}\label{sec:conclusions}
Parton fragmentation functions are a fundamental component of modern high-energy physics, providing the indispensable bridge between the colored world of quarks and gluons and the experimentally observed world of colorless hadrons. From their phenomenological origins, the concept of FFs has evolved into a cornerstone of QCD, supported by a rigorous theoretical framework based on factorization and DGLAP evolution. Global analyses combining world data from $e^+e^-$, SIDIS, and $pp$ collisions have determined the unpolarized FFs for light hadrons with remarkable precision, now reaching NNLO accuracy.

Despite this progress, significant challenges and exciting frontiers remain. A key area of research is the exploration of the three-dimensional structure of hadronization. This requires a comprehensive understanding of spin- and transverse-momentum-dependent fragmentation functions, which connect spin and momentum correlations within QCD. While the collinear polarized FFs describing spin transfer can be obtained by integrating TMD-FFs, mapping the full 3D picture encoded in functions like the Collins function is a central focus of current and future research. Dihadron FFs provide a powerful alternative route to this 3D structure. Furthermore, understanding how the fragmentation process is modified in the presence of a nuclear medium, both cold and hot, remains a crucial goal in nuclear physics, essential for characterizing the quark-gluon plasma.

Future research will be driven by synergies between experiment, theory, and first-principles calculations. The future EIC will provide unprecedented precision and kinematic coverage for SIDIS, revolutionizing our knowledge of spin-dependent and multi-dimensional FFs. Continued high-statistics measurements of jet substructure and heavy flavor fragmentation at the LHC will further constrain gluon and heavy quark FFs. On the theoretical side, pushing perturbative calculations to higher orders, developing a unified framework for TMD and collinear physics, and improving models of hadronization and medium modification are all crucial. While first-principles lattice QCD calculations of FFs remain a challenge due to their real-time nature, new theoretical and computational approaches may provide future opportunities. The ongoing synergy across these domains ensures that the study of parton fragmentation will continue to be a vibrant and fruitful area of research for years to come.

\begin{ack}[Acknowledgments]%
 We thank NPC members for valuable discussions. 
 \end{ack}

\bibliographystyle{Numbered-Style} 
\bibliography{reference}

\end{document}